# Photoconductive nonpolar liquids based on azobenzene

**Carlo Rigoni[1*], Promeet K. Saha[1], Jinyu Sheng[1,2], Rafal Klajn[1*]**

[1] Institute of Science and Technology Austria, Am Campus 1, 3400 Klosterneuburg, Austria

[2] College of Chemistry, Chemical Engineering and Materials Science, Soochow University, Suzhou, Jiangsu 215123, China

* Corresponding author; E-mail: carlo.rigoni@ist.ac.at, rafal.klajn@ist.ac.at

## Abstract

**The weakly conductive properties of mixtures of organic surfactants in nonpolar liquids are fundamental to many electrohydrodynamic phenomena and underpin several cutting-edge technologies, particularly the development of electrophoretic displays. To date, tuning the electrical properties of these systems has involved modifying their composition, including surfactant type and concentration, water content, and the carrier liquid, all of which influence their behavior. Here, we use photoresponsive molecules to control electric phenomena in nonpolar liquids externally with light irradiation, thereby rendering them photoconductive. In particular, we examine azobenzene solutions in toluene, whose conductivity can be adjusted by two colors of light: UV induces *trans→cis* isomerization, leading to an increase in conductivity, while blue light triggers *cis→trans* back-isomerization, decreasing conductivity. The findings of this study suggest new ways to expand the applications of weakly conductive organic fluids, such as in self-regulating devices that respond to sunlight or in externally programmable displays.**

## Introduction

Nonpolar liquids are known to be effective insulators. When measurable, the current flow in these systems is small because it is difficult to generate and sustain free charged species in nonpolar media. Introducing surfactants as charge-control agents (CCAs) overcomes this limitation by increasing conductivity ($\sigma$) by several orders of magnitude[1,2], giving rise to a family of fluids with leaky dielectric properties. Owing to these desirable properties—which enable, for example, the displacement of charged particles within the fluid with very low power consumption—these fluids have found use in various technological applications, such as electrophoretic or electronic ink displays[3,4] (present in e-readers, smart watches, paper tablets, and switchable billboards). Additionally, leaky dielectric fluids have been essential in studying numerous electrohydrodynamic phenomena[5,6], including electrically driven solid particle charging, rotation, and translation[7–11], as well as three-dimensional (3D) liquid droplet rotation and deformation[6,12–14], advancing research in active matter[8,11], microfluidics[14,15], and multiresponsive materials[10]. Despite the proven success of CCAs in increasing conductivity in nonpolar liquids, the ability to tune the electrical properties of the system stops at the sample preparation stage: once the concentration of CCAs and trace impurities (water or ions) in the fluid mixture is established, the fluid's responsiveness to electric fields becomes fixed.

Photoresponsive molecules that can change their electrical properties when exposed to light offer a promising way to introduce tunability after sample preparation. One class of molecules that meets these criteria is azobenzenes[16]. When exposed to light, azobenzenes undergo *trans→cis* isomerization, which involves (i) a substantial change in molecular geometry (from planar to bent); and (ii) the development of an electric dipole moment $\mathbf{p}$. Azobenzene derivatives have already proven effective in tuning the electrical properties of soft materials[17–20] and even the conductivity of emulsions[21]. In fact, although it has long been established that photoresponsive molecules can be used to adjust the conductivity of polymer-based soft materials[22–24] on a macroscopic scale, their ability to modulate the electrical properties of nonpolar fluids has received little attention[25].

Here, we reversibly switch the conductivity of an azobenzene-based leaky dielectric fluid using light irradiation. First, we describe the approach used to perform the experiments and assess the effect of light irradiation. Next, we present the results obtained by controlling the light stimulus and adjusting the mixture's responsiveness. Finally, we examine how the photoconductivity evolves under repeated stimuli and evaluate the system's performance over time. We demonstrate that we can modulate the fluid's conductivity with different colors of light, tune the magnitude of the changes by varying either the intensity or duration of irradiation, and sustain the system's responsiveness across many cycles. The high modulation of the conductive properties of nonpolar fluids achieved in this study suggests potential for future technological applications.

## Experimental system and photoconductive behavior

The photoconductive materials used in this study are toluene solutions of azobenzenes functionalized with alkoxy groups at the *para* (i.e., 4 and 4') positions. Specifically, we focus on 4,4'-di-*n*-octyloxyazobenzene (L-C8Azo, where 'L' stands for 'linear'). Upon absorbing a UV photon (365 nm; we worked with a multi-LED CoolLED pE-4000 system; see Methods), *trans*-L-C8Azo is promoted to a photoexcited state (PES) that can either relax back to the initial, planar *trans* isomer, or convert into the metastable, bent *cis* form (Fig. 1a)[26], with a significant dipole moment typical of 4,4'-disubstituted *cis*-azobenzenes (~4.4 D)[27]. The *cis*→*trans* isomerization can occur either through slow thermal relaxation or rapidly, by photoexcitation with blue light (460 nm). The progress of this interconversion can be followed by monitoring changes in the UV-vis absorption profiles (Supplementary Fig. 1). Electric fields were applied to the photoconductive solutions using Hele–Shaw microelectrode cells, composed of two ITO-coated glass slides with an active area of 1 cm$^2$, separated by 10 μm (Fig. 1b and Supplementary Fig. 2). Applying a voltage drop $U$ across the cell generates a uniform electric field $E = U/h$, driving a Faradaic current through the cell.

Following a short transient peak in current due to the capacitive nature of the cell, the current stabilizes at the steady state (Fig. 1c, $t_1$) determined by the balance between the charge-exchange rate of the molecules at the electrodes and electrophoresis. The current of the L-C8Azo solution in toluene is significantly higher than that of pure toluene under the same conditions; therefore, L-C8Azo acts as the primary charge carrier in the mixture. Since L-C8Azo is electron-rich, the most likely charging mechanism involves losing an electron at the positive electrode, with the resulting electron hole localized mainly at the center of the molecule. This charged molecule is then transported electrophoretically to the negative electrode, where its positive charge is neutralized.

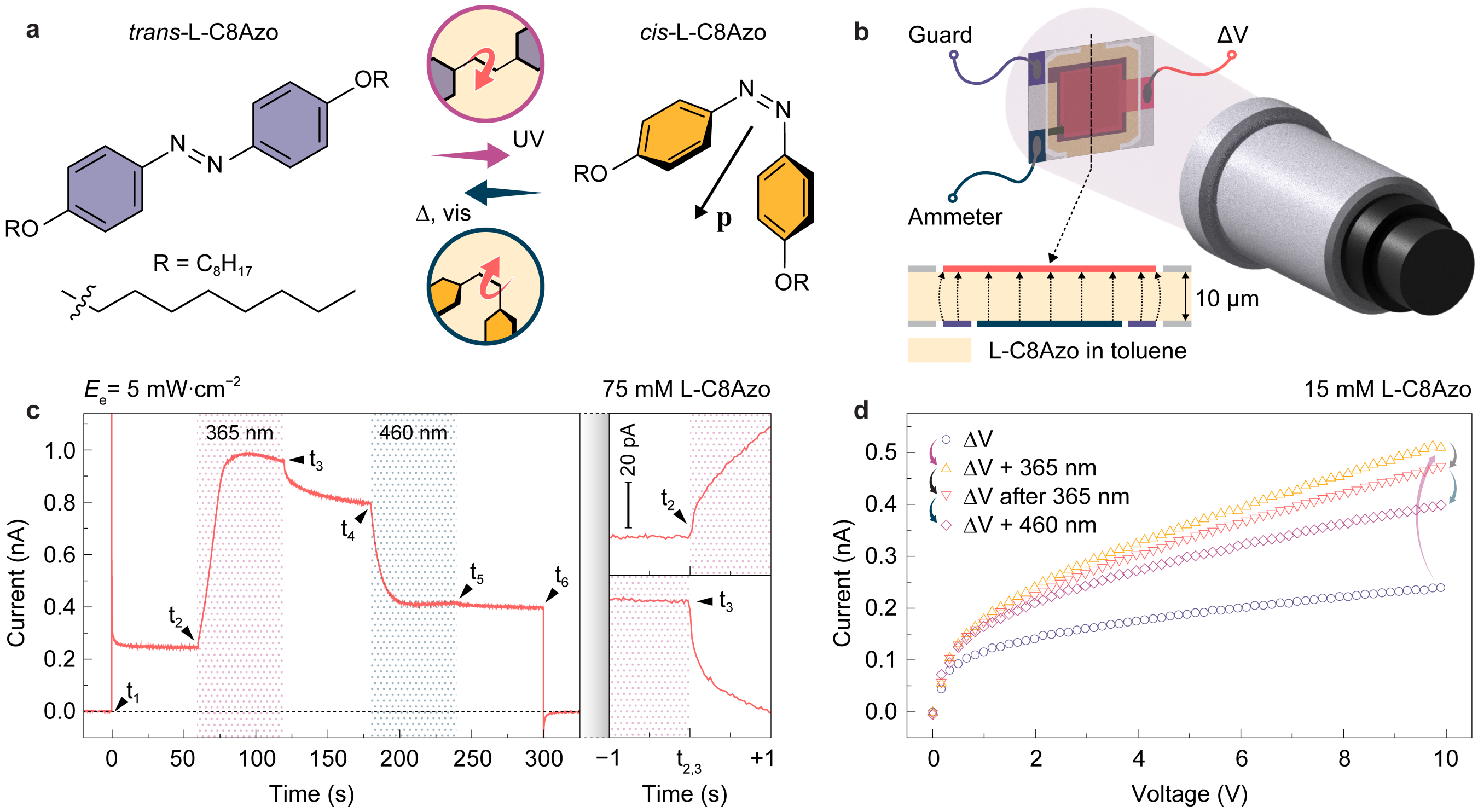

**Fig. 1 | Photoresponsiveness of the current in solutions of di-*para*-substituted azobenzenes in nonpolar liquids. a,** The light-induced transformation between the *trans* and *cis* isomers of L-C8Azo (4,4'-dioctyloxyazobenzene). **b,** Barebone schematic of the experimental setup illustrating the dimensions of the microelectrode cell and the LED used for irradiation. **c,** A representative current measurement conducted to evaluate photoresponsivity. The current can be modulated by external light irradiation. **d,** Quasi-static cyclic voltammograms of a 15 mM solution of L-C8Azo in toluene at different *trans*/*cis* ratios.

Upon UV irradiation ($t_2$), the current rises due to the higher charge exchange rates of the photogenerated species (i.e., molecules in the PES and *cis*-azobenzene) compared to *trans*[28–30]. We attribute the initial sharp step (occurring on a timescale faster than the measurement rate of 40 Hz; see the side plot in Fig. 1c) to (i) the increased polarizability and (ii) the lower ionization potential of L-C8Azo in the PES[28]. Together, these factors enhance the charge exchange rate at the electrodes, leading to higher conductivity (see Supplementary Fig. 3 for additional support of this interpretation). After this initial step, the current rises exponentially, reflecting the kinetics of *trans→cis* isomerization (see Supplementary Note 1). During this period, the gradual formation of *cis*-L-C8Azo drives the current increase, as its high dipole moment promotes a stronger directional interaction—and thus, more effective charge exchange—with the conductive surfaces of the charged electrodes than *trans*-L-C8Azo. Additionally, the higher energy levels of electrons in *cis*-L-C8Azo help facilitate tunneling at the electrodes[31]. The current reaches a maximum value after prolonged irradiation, once a *cis*-rich photostationary state (PSS) is established (typically with less than 5 mol% residual *trans*-L-C8Azo; see Supplementary Note 2). As a result of photoirradiation, the fluid's temperature also rises, which further increases conductivity.

Switching off the light source ($t_3$) causes a sharp step-like decrease in current due to depopulation of the PES, followed by a gradual decline as thermal equilibrium is established with the surrounding glass in the microelectrode cell, until a new steady state is reached. Subsequent blue light irradiation ($t_4$) shifts the population of L-C8Azo molecules back toward the *trans* state, thereby decreasing the current in the cell. Note that the molar extinction coefficient of a typical *cis*-azobenzene at 460 nm is an order of magnitude lower than that of *trans*-azobenzene at 365 nm (Supplementary Fig. 1); therefore, the population of molecules in a PES following irradiation with 460 nm light ($t_4$) is lower, and their contribution to the current is comparable to the measurement noise. Since the new PSS contains some residual *cis*-L-C8Azo (Supplementary Fig. 1), the current does not return to its initial value even after prolonged blue light irradiation.

Upon switching off the blue light ($t_5$), the current remains stable until the voltage is turned off ($t_6$). Besides temperature equilibration, the only other factor that might influence the current at this point is the thermal *cis→trans* isomerization of L-C8Azo, which occurs on a timescale significantly longer than the duration of our measurements, even with a voltage applied (Supplementary Fig. 4), and thus has minimal impact. This dependence of current on light irradiation can be seen, albeit with varying degrees of magnitude, across a wide range of voltages (Fig. 1d). The current–voltage profile is again determined by the voltage dependencies of (i) the charging rate of the molecules at the electrode[31] (which also differs between the *trans* and *cis* isomers) and (ii) electrophoresis (see Supplementary Note 1).

## Tuning the stimulus

The conductivity of L-C8Azo solutions can be regulated by independently adjusting the irradiation duration or intensity. Light intensity has a complex impact on the current profile. As shown in Fig. 2b, increasing light intensity raises both the initial slope (due to a higher availability of photons for the *trans*→*cis* photoisomerization; Fig. 2b) and the maximum current value obtained. Since changing the light intensity does not alter the fraction of *cis*-L-C8Azo in the PSS, the increase in maximum current partly results from a higher concentration of molecules in the PES (which is linearly proportional to the irradiation intensity; Fig. 2c). Furthermore, a higher light intensity leads to higher final steady-state temperatures—determined by the interplay between the photothermal heating rate and the rate of heat transfer from the fluid to the surrounding glass—and higher steady-state currents. Due to these factors, the current in the PSS ($I_{PSS}$) depends linearly on the irradiation intensity (Fig. 2d). The only notable exception to this trend is the value at the lowest irradiation intensity, where the experimental timeframe is too short to reach a PSS, leading to a lower final current. The contribution of the PESs to the PSS current can also be evaluated by analyzing the sharp current drop that occurs immediately after irradiation is stopped ($t$ = 300 s). Again, we find a linear dependence of the magnitude of current decrease on light intensity (Fig. 2e). The role of temperature increase (compared to the PSS-associated drop) is further confirmed by changes in current after irradiation is turned off (Supplementary Fig. 5). However, if stopping the 365 nm irradiation is accompanied by turning on the 460 nm LED, we again observe that the current changes are primarily driven by azobenzene photoisomerization and photothermal effects (Supplementary Fig. 6).

The dependence of current on the relative populations of *cis*- and *trans*-L-C8Azo is also evident from how the current profiles change over irradiation time (Fig. 3). Irradiating the sample at λ = 365 nm for longer periods (with constant light intensity) results in higher final current values. This dependence persists

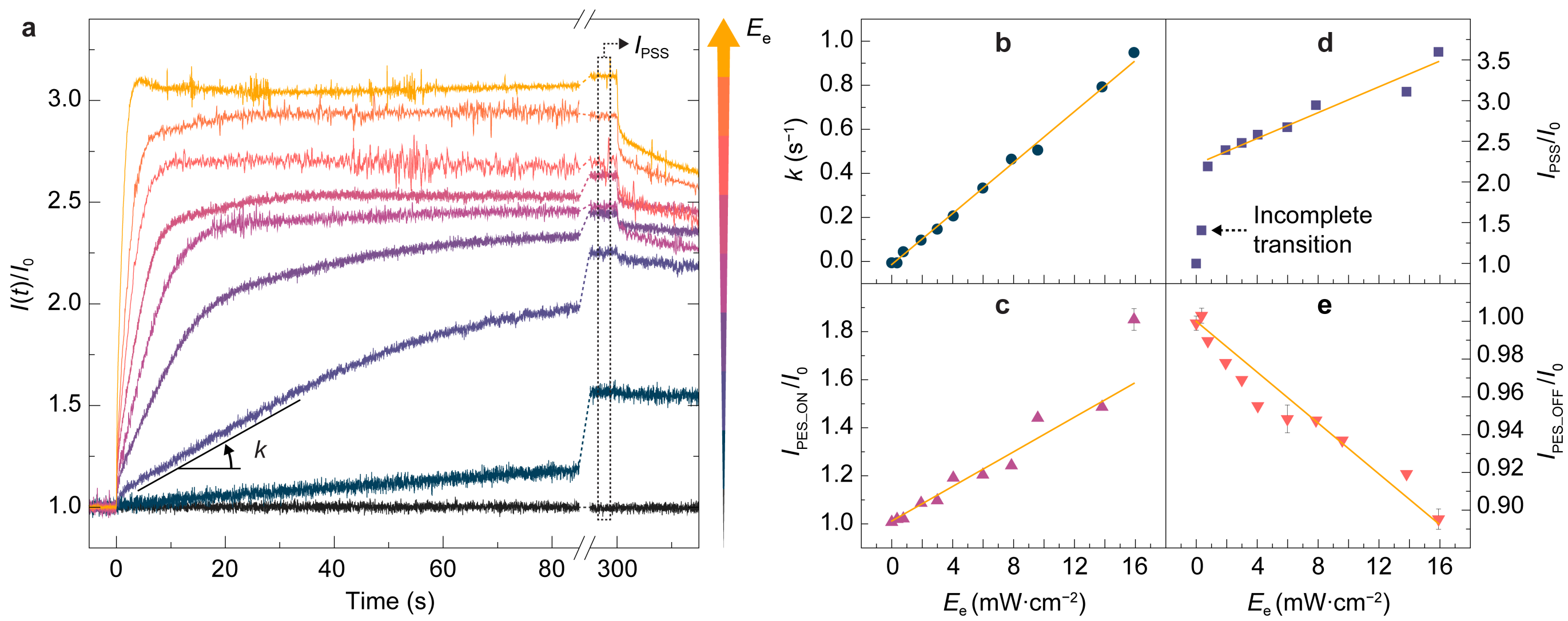

Relative increase in current within the cell upon 365 nm light irradiation at different intensities. **b,** Rate of *trans*→*cis* isomerization, derived from the initial slope of the current measurement data. **c,** Initial jump in the current caused by populating the photoexcited state (PES). **d,** Relative increase in current at $t$ = 200 s compared to immediately before photoirradiation. **e,** Sudden drop in current after stopping the irradiation indicates the disappearance of the PES. The applied voltage is 2 V for all measurements.

even after irradiation stops (i.e., as the contribution of the PES disappears and the current gradually reaches a steady state; Fig. 3a). The effect of temperature increase can be quantified by evaluating the current decay after UV irradiation is turned off (Supplementary Fig. 7). The current's dependence on *cis*/*trans* isomerization is further evident from the decrease in current during subsequent irradiation with 460 nm light (Fig. 3b). In both cases (i.e., irradiation with λ = 365 nm or 460 nm), the current changes exponentially with irradiation time (see the insets in Fig. 3a,b), reflecting the forward and backward isomerization reactions. Analogous observations were made when decoupling the irradiation from the voltage application (Supplementary Fig. 8).

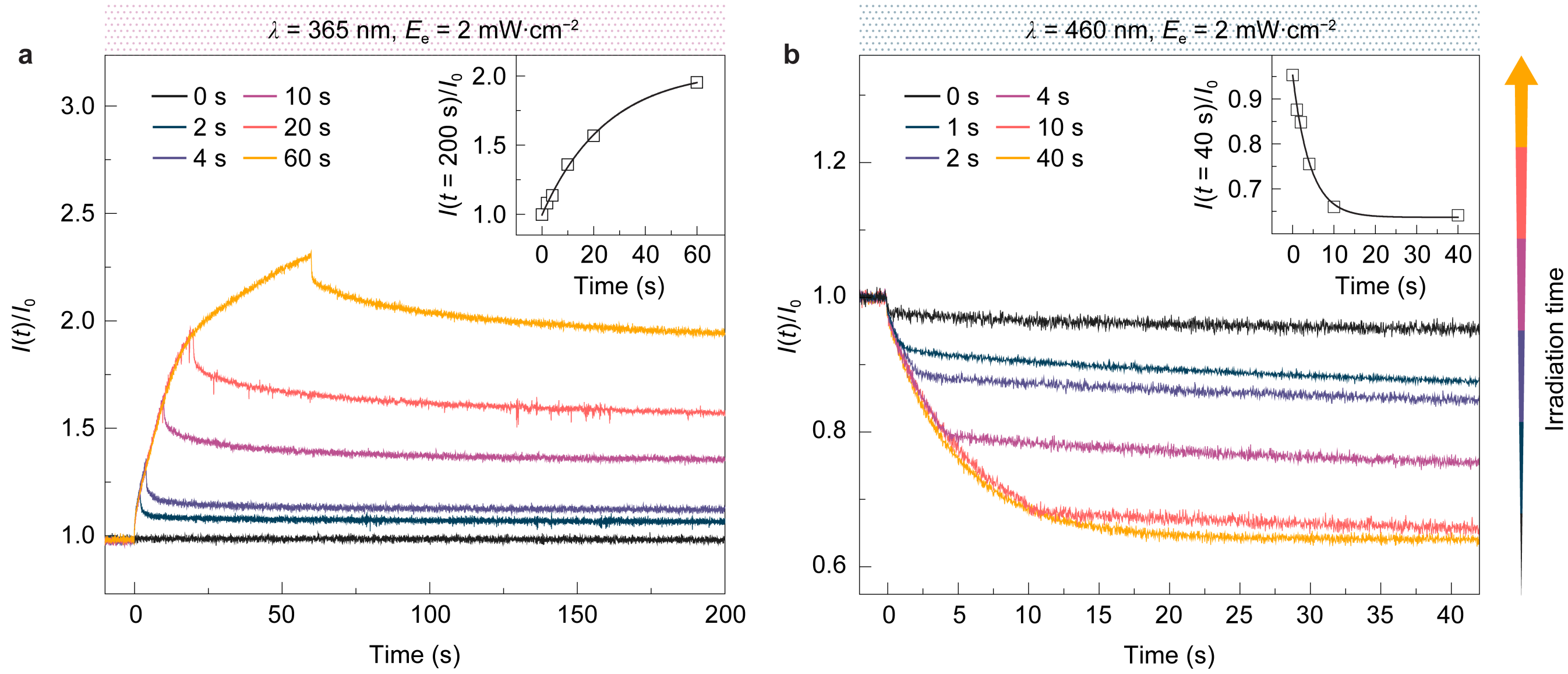

**Fig. 3 | Response of a 15 mM toluene solution of L-C8Azo to various irradiation times. a,** Relative increase in current in the cell upon light irradiation for different durations with $\lambda$ = 365 nm at $E_e$ = 10 mW·cm$^{-2}$. Inset: relative increase in current measured at $t$ = 200 s. **b,** Relative decrease in current due to irradiation with $\lambda$ = 460 nm at $E_e$ = 10 mW·cm$^{-2}$ for different durations. The data in the inset show the relative decrease in current 40 s after the start of irradiation. The applied voltage was 2 V for all measurements.

## Tuning the responsivity

The responsiveness of the studied fluids can be tuned using two approaches without drastically changing their working principle: varying the concentration of the compound or changing the substituents at the *para* positions of the azobenzene core. The high solubility of L-C8Azo in toluene allows us to study photoconductivity across a broad concentration range. Current measurements at increasing L-C8Azo concentrations (Fig. 4a,b and Supplementary Fig. 9) reveal a conductivity increase of over two orders of magnitude at the highest concentration (75 mM) compared to pure toluene. At concentrations ≥25 mM, the increase in current upon UV irradiation deviates from the expected exponential behavior (Supplementary Note 1). This discrepancy, particularly evident at 75 mM, can be attributed in part to the limited light penetration depth: layers closer to the LED source absorb light preferentially, screening the lower layers and hindering isomerization in that region of the cell. However, this inner-filter effect alone cannot fully explain the nonlinear increase in conductivity that follows a power-law curve (Fig. 4b). We hypothesize that this unexpectedly large increase originates from intermolecular interactions involving a PES, which can lead to charge disproportionation[32]; such interactions are more likely at high azobenzene concentrations and may boost solution conductivity.

We hypothesized that conductivity can also be affected by changing the alkyl substituent at the *para* positions of the azobenzene core. To this end, we synthesized and studied three additional 4,4'-disubstituted azobenzenes, which carry shorter (C1Azo), longer (L-C12Azo), or branched (B-C8Azo) alkyl chains (Fig. 4c and Supplementary Note 2). Working with 10 mM solutions of B-C8Azo, we observed a ~35-fold increase in conductivity compared to pure toluene in the dark, and a ~60-fold increase under UV irradiation – i.e., over twice the conductivity of L-C8Azo. In contrast, the conductivities of L-C12Azo solutions were only ~2- and ~5-fold higher than those of pure toluene before and after irradiation, respectively. These differences could not be attributed to different photoisomerization behaviors – in fact, all four azobenzene derivatives exhibited similar photoisomerization kinetics and PSS compositions (Supplementary Fig. 10). We note, however, that although the solubility of all four compounds in toluene is well above 10 mM (and thus all have similar mobilities), B-C8Azo's branched alkyl substituents most likely increase the solubility of the more polar *cis* isomer as well as charged species and molecules in the PES, all of which are expected to be solvated less efficiently in the nonpolar medium, leading to aggregation and lower mobilities. In contrast, the poorer solvation and better packing of the linear chains of L-C8Azo and L-C12Azo facilitate aggregation, thereby decreasing charge mobility. Finally, the methoxy substituents have the least steric bulk, making C1Azo the most prone to aggregation. Overall, the different behaviors of differently substituted azobenzenes can be attributed to the interplay of solvation, intermolecular forces, and steric effects.

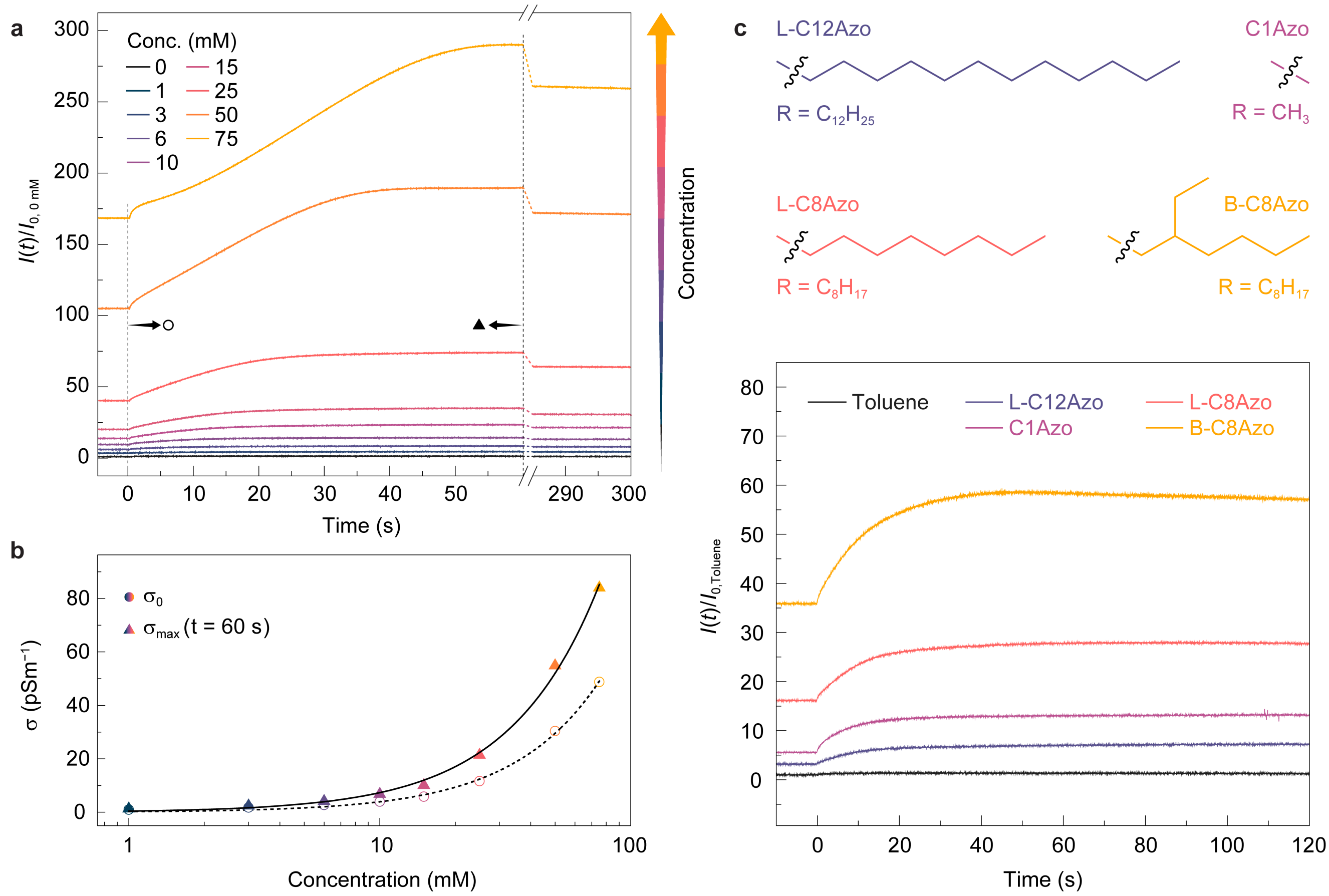

**Fig. 4 | Photoconductivity as a function of the concentration and the *para* substituent identity. a,** Time–current plots for different concentrations of L-C8Azo in toluene (applied voltage, $U$ = 2 V, irradiation wavelength, $\lambda$ = 365 nm, irradiation intensity, $E_e$ = 10 mW·cm$^{-2}$; irradiation time, $t$ = 300 s). **b,** Conductivity as a function of the concentration before irradiation ($\sigma_0$) and after irradiation for $t$ = 60 s (to ensure that the highest conductivity value, $\sigma_{max}$, has been reached; see panel **a**). Fitting with the power law function $\sigma(c) = \alpha c^{\beta}$ (where $c$ is the concentration, and $\alpha$ and $\beta$ are fitting parameters ($\alpha_{max} \approx 2\alpha_0$ and $\beta_{max} \approx \beta_0 \approx 1.25$) suggests that increasing concentration has a smaller than expected effect on photoconductivity. **c,** Time–current plots for 10 mM solutions of azobenzenes with different *para* substituents ($U$ = 2 V, $\lambda$ = 365 nm, $E_e$ = 10 mW·cm$^{-2}$, and $t$ = 300 s).

## Performance

Finally, we studied the performance of L-C8Azo solutions over multiple irradiation cycles to assess their potential for future applications (such as a light-powered switch). First, we repeated the measurement of L-C8Azo (Fig. 1c) properties several times within the same microelectrode cell to evaluate the level of degradation caused by photoirradiation and voltage application (Fig. 5d–f). We found that repeating the measurement multiple times leads to an overall increase in conductivity (Fig. 5a), but this increase heavily depends on experimental conditions. Avoiding irradiation of the sample removes this effect completely (Fig. 5e, blue circles), and the same result can be achieved by excluding oxygen (sample preparation under an argon atmosphere) (yellow triangles in Fig. 5b; see also Supplementary Fig. 11). Photodegradation can also be reduced by decoupling irradiation from voltage application (Fig. 5b, purple triangles). Therefore, we can conclude that i) the process is dependent on the presence of oxygen, and ii) the increase in conductivity over time is most probably linked to a steady photodegradation of the sample that occurs when both a voltage and irradiation are applied simultaneously (in the presence of oxygen). Furthermore, by comparing the ratios between the value of the current at the PSS ($I_{\mathrm{PSS}}$) to that before starting irradiation ($I_0$), we can evaluate whether photodegradation impacts the photoresponsivity of the mixture due to *trans→cis* isomerization of L-C8azo (Fig. 5c). After the initial loss of responsivity (δ) due to inefficient photoswitching (Supplementary Fig. 1), the $I_{\mathrm{PSS}}/I_0$ values decay exponentially to a steady-state value ($\delta'_\infty \approx 2$) within a small number of cycles (~3.5). These findings suggest that photodegradation minimally affects the switching properties of L-C8Azo in toluene.

An additional proof that photoresponsiveness is not significantly affected by photodegradation can be obtained by examining the current during repeated cycles of i) irradiation at 365 nm, ii) irradiation at 460 nm, and iii) dark incubation, testing the system as a light-powered current switch (Fig. 5d–f). By monitoring the time evolution of the local maxima and minima for each cycle (Fig. 5e) and their differences (Fig. 5f), we observe that although the current gradually increases due to photodegradation, the value of $\Delta I/I_0$—a measure of the responsivity of the photoconductive mixture—stabilizes to a steady-state value ($\delta'_\infty \sim 3.5$) within a short time ($\tau_\delta \sim 3$ min).

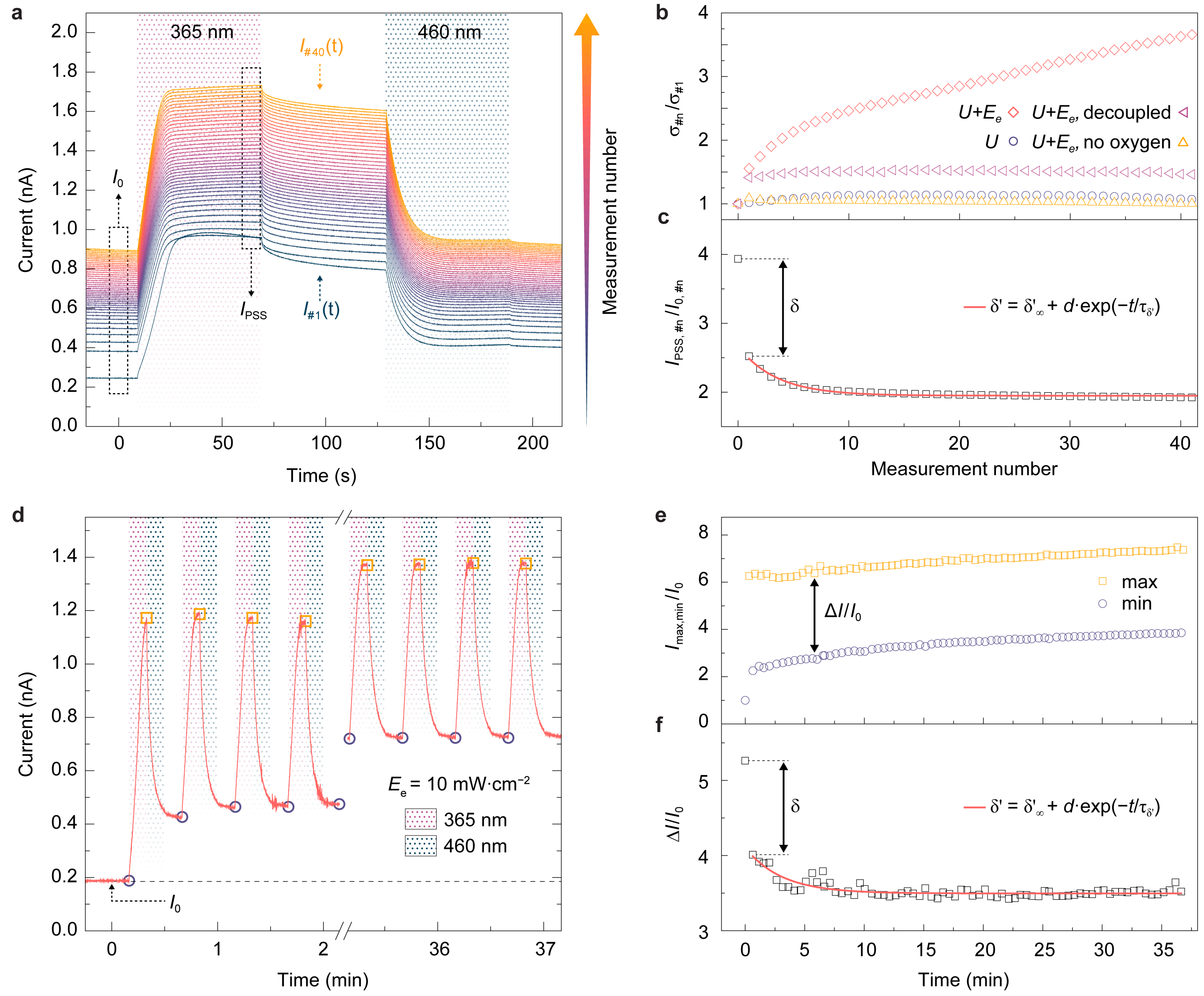

**Fig. 5 | Photodegradation and responsivity changes in L-C8Azo solutions under repeated stimulation. a,** Current data from over 40 consecutive measurements with a 15-minute pause between each. **b,** Evolution of conductivity during consecutive runs in the presence of voltage and irradiation ($U + E_e$), voltage only ($U$), voltage and irradiation applied separately ($U + E_e$, decoupled), and in the presence of voltage and irradiation but in the absence of oxygen ($U + E_e$, no oxygen). **c,** Relative increase of the photostationary state current ($I_{PSS}$) during consecutive measurements. **d,** Time evolution of the current during alternating irradiation with 365 nm and 460 nm at $E_e$ = 10 mW·cm$^{-2}$. **e,** Time evolution of the local maxima (yellow squares) and minima (violet circles) of the current in panel **d**. **f,** Relative difference between the values of the local maxima and minima over time.

## Conclusions

Our work demonstrates that nonpolar fluids can be endowed with photoconductivity. We show that the electrical conductivity of solutions of *para*-substituted azobenzenes in toluene can be readily tuned by exposure to UV and visible light. These conductivity changes are primarily due to the reversible photoisomerization between *trans*- and *cis*-azobenzene, which have different charge exchange rates, with minor contributions from other factors, such as photothermal effects and the existence of molecules in the photoexcited state. The photoconductivity correlates with the concentration of the photoresponsive compound in the solution. Changing the substituent group at the *para* position affects the solution's photoconductivity, demonstrating another way to control the system's behavior. Lastly, photodegradation caused by repeated cycling has only a small effect on the photoconductive properties and can be effectively minimized by removing oxygen from the cell. The simplicity of our design provides the basis for engineering more complex and diverse photoconductive devices. For example, modifying the other positions of the azobenzene core could lead to molecules that switch under visible light, as in the case of tetra-*ortho*-substituted azobenzenes, whose *cis*-isomer has a half-life of several hundred days, or compounds with *cis* half-lives in the milliseconds regime, such as *ortho*-hydroxy derivatives[26]. The diversity of switching behaviors within the azobenzene family ensures that photoconductive liquids based on these molecules can meet different needs in technological applications.

## Methods

***Microelectrode cell preparation:*** The parallel-face Hele–Shaw microelectrode cells (Instec SG100T100uNOPI, distance between plates 10 µm, active area 1 $cm^2$) were connected to metal wires using a specific alloy of solder (Cerasolzer GS182) with an ultrasonic soldering iron (MBR Electronics USS-9210 Ultrasonic Soldering System). The cell was filled with a micropipette (Eppendorf Research Plus) and sealed with a high-viscosity fluorinated paste (Krytox® XHT-BDZ grease) to limit evaporation during measurements and prevent sample contamination.

***Current measurements:*** Voltage was applied to microelectrode cells enclosed in a custom-made Faraday cage (Supplementary Fig. 2) using an Arb Waveform Generator (Multicomp Pro MP750288). In the experiments, voltage was consistently applied as a DC 2 V step, except for the data shown in Fig. 1d and Supplementary Note 1, where a triangular wave was used instead (for Fig. 1d, the wave had a period of 300 s). The current flowing in the cell was measured with a femto/picoammeter (Keysight B2981B). In all experiments, current data were collected at a sampling rate of 40 Hz and with a 0.02 s aperture to minimize noise from the 50 Hz power line frequency. For different experimental protocols, both the waveform generator and the ammeter were controlled using Python scripts. All measurements were performed on freshly prepared solutions to improve data reproducibility. Typically, when using microelectrode cells from the same batch (INSTEC) and freshly prepared photoconductor samples, the coefficient of variation (standard deviation divided by the mean) of the absolute current values was less than 5%. However, using microelectrode cells from different batches significantly affected the reproducibility of absolute values; therefore, all measurements comparing data to assess the effect of changing a specific parameter (e.g., light intensity, irradiation time, concentration) were performed with the same batch of microelectrode cells to prevent errors.

***Light irradiation:*** Microelectrode cells filled with different samples were irradiated *in situ* during current measurements using a multi-LED source (CoolLED, pE-4000) with a collimator (CoolLED, pE-Universal Collimator) positioned at a distance of 5 cm from the cell. For all photoirradiation experiments, the LED source was fitted with one of two filters: a bandpass filter centered at 365 nm with a 10 nm spectral window (ET365/10x, AHF analysentechnik) for UV irradiation (pE-4000's 365 nm LED), and a bandpass filter centered at 470 nm with a 40 nm window (ET470/40x, AHF analysentechnik) for blue light irradiation (pE-4000's 460 nm LED). The LED control was integrated into the same custom-made Python scripts used for current measurements. The irradiance was measured with a thermal power meter (Thorlabs S310C, Thorlabs PM100D).

***UV-vis absorption measurements:*** Absorption spectra of azobenzene solutions were recorded using a UV-vis absorption spectrophotometer (Cary 60, Agilent). Spectra were collected over a range from 200 nm to 800 nm, sampling every 0.5 nm. The collection time for each profile was 120 s. All measurements were performed in a semi-micro quartz cuvette (117-104-10-40 Hellma Analytics) with a 10 mm path length at a concentration of 15 μM for all samples, except for the data collected for Supplementary Fig. 4, where the measurements were performed directly in microelectrode cells with a path length of 10 μm and a concentration of 15 mM.

## Supplementary Materials

Supplementary Figures 1–22

Supplementary Notes 1, 2

## References

1 Smith, G. N. & Eastoe, J. Controlling colloid charge in nonpolar liquids with surfactants. *Phys. Chem. Chem. Phys.* **15**, 424–439 (2013).

2 Strubbe, F. & Neyts, K. Charge transport by inverse micelles in non-polar media. *J. Phys. Condens. Matter* **29**, 453003 (2017).

3 Rogers, J. A. *et al.* Paper-like electronic displays: Large-area rubber-stamped plastic sheets of electronics and microencapsulated electrophoretic inks. *Proc. Natl. Acad. Sci. U.S.A.* **98**, 4835–4840 (2000).

4 Eshkalak, S. K. *et al.* Overview of electronic ink and methods of production for use in electronic displays. *Opt. Laser Technol.* **117**, 38–51 (2019).

5 Saville, D. A. Electrohydrodynamics: the Taylor-Melcher leaky dielectric model. *Annu. Rev. Fluid Mech.* **29**, 27–64 (1997).

6 Vlahovska, P. M. Electrohydrodynamics of drops and vesicles. *Annu. Rev. Fluid Mech.* **51**, 305–330 (2019).

7 Comiskey, B., Albert, J. D., Yoshizawa, H. & Jacobson, J. An electrophoretic ink for all-printed reflective electronic displays. *Nature* **394**, 253–255 (1998).

8 Bricard, A., Caussin, J. B., Desreumaux, N., Dauchot, O. & Bartolo, D. Emergence of macroscopic directed motion in populations of motile colloids. *Nature* **503**, 95–98 (2013).

9 Morin, A., Desreumaux, N., Caussin, J.-B. & Bartolo, D. Distortion and destruction of colloidal flocks in disordered environments. *Nat. Phys.* **13**, 63–67 (2017).

10 Cherian, T., Sohrabi, F., Rigoni, C., Ikkala, O. & Timonen, J. V. I. Electroferrofluids with nonequilibrium voltage-controlled magnetism, diffuse interfaces, and patterns. *Sci. Adv.* **7**, eabi8990 (2021).

11 Garza, R. R., Kyriakopoulos, N., Cenev, Z. M., Rigoni, C. & Timonen, J. V. I. Magnetic Quincke rollers with tunable single-particle dynamics and collective states. *Sci. Adv.* **9**, eadh2522 (2023).

12 Salipante, P. F. & Vlahovska, P. M. Vesicle deformation in DC electric pulses. *Soft Matter* **10**, 3386–3393 (2014).

13 Brosseau, Q. & Vlahovska, P. M. Streaming from the equator of a drop in an external electric field. *Phys. Rev. Lett.* **119**, 034501 (2017).

14 Raju, G., Kyriakopoulos, N. & Timonen, J. V. Diversity of non-equilibrium patterns and emergence of activity in confined electrohydrodynamically driven liquids. *Sci. Adv.* **7**, eabh1642 (2021).

15 Dong, Q. & Sau, A. Breakup of a leaky dielectric drop in a uniform electric field. *Phys. Rev. E* **99**, 043106 (2019).

16 Jerca, F. A., Jerca, V. V. & Hoogenboom, R. Advances and opportunities in the exciting world of azobenzenes. *Nat. Rev. Chem.* **6**, 51–69 (2022).

17 Zhang, S., Liu, S., Zhang, Q. & Deng, Y. Solvent-dependent photoresponsive conductivity of azobenzene-appended ionic liquids. *Chem. Commun.* **47**, 6641–6643 (2011).

18 Li, Z. *et al.* A reversible conductivity modulation of azobenzene-based ionic liquids in aqueous solutions using UV/vis light. *Phys. Chem. Chem. Phys.* **20**, 12808–12816 (2018).

19 Nishikawa, H., Sano, K. & Araoka, F. Anisotropic fluid with phototunable dielectric permittivity. *Nat. Commun.* **13**, 1142 (2022).

20 Nguyen, P. H. *et al.* Reversible modulation of conductivity in azobenzene polyelectrolytes using light. *ACS Appl. Polym. Mater.* **5**, 4698–4703 (2023).

21 Bufe, M. & Wolff, T. Reversible switching of electrical conductivity in an AOT−isooctane−water microemulsion via photoisomerization of azobenzene. *Langmuir* **25**, 7927–7931 (2009).

22 Kawai, T., Nakashima, Y. & Irie, M. A novel photoresponsive π-conjugated polymer based on diarylethene and its photoswitching effect in electrical conductivity. *Adv. Mat.* **17**, 309–314 (2005).

23 Chang, C.-W. *et al.* Developing stretchable and photo-responsive conductive films by incorporation of spiropyran into poly(ionic liquid)s. *ACS Appl. Polym. Mater.* **5**, 4210–4216 (2023).

24 Chen, S. *et al.* Photo responsive electron and proton conductivity within a hydrogen-bonded organic framework. *Angew. Chem. Int. Ed.* **62**, e202308418 (2023).

25 Yoon, C.-M., Jang, Y., Noh, J., Kim, J. & Jang, J. Smart fluid system dually responsive to light and electric fields: An electrophotorheological fluid. *ACS Nano* **11**, 9789–9801 (2017).

26 Bandara, H. D. & Burdette, S. C. Photoisomerization in different classes of azobenzene. *Chem. Soc. Rev.* **41**, 1809–1825 (2012).

27 Tong, X., Wang, G., Soldera, A. & Zhao, Y. How can azobenzene block copolymer vesicles be dissociated and reformed by light? *J. Phys. Chem. B.* **109**, 20281–20287 (2005).

28 Gerischer, H. & Willig, F. Reaction of excited dye molecules at electrodes. *Top. Curr. Chem.* **61**, 31–84 (1976).

29 Casadei, T. *et al.* Rapid electrochemical assessment of excited-state quenching dynamics. *ACS Catal.* **15**, 16938–16952 (2025).

30 Lee, Y.-M., Nam, W. & Fukuzumi, S. Redox catalysis via photoinduced electron transfer. *Chem. Sci.* **14**, 4205–4218 (2023).

31 Bevan, K. H., Foong, Y. W., Shirani, J., Yuan, S. & Abi Farraj, S. Physics applied to electrochemistry: Tunneling reactions. *J. Appl. Phys.* **129**, 090901 (2021).

32 Schatz, D. & Wegner, H. A. Electrochemistry of azobenzenes and its potential for energy storage. *J. Org. Chem.* **90**, 5336–5342 (2025).

## Acknowledgements

This work was supported by the Austrian Science Fund (FWF) through project no. ESP1138624 (C.R.) and the Academy of Finland through project no. 342038 (C.R.).

## Author Contributions

C.R. led the project, prepared samples for measurements, performed electrical and spectroscopic measurements, analyzed the data, prepared the main and supplementary figures, and wrote the first draft of the manuscript. P.S. and J.S. synthesized the azobenzene derivatives and performed the NMR measurements. C.R., P.S., and R.K. curated the interpretation of the results. R.K. supervised J.S. and P.S.'s work and edited the manuscript, with input from all authors.

# Supplementary Information

# Photoconductive nonpolar liquids based on azobenzene

**Carlo Rigoni[1*], Promeet K. Saha[1], Jinyu Sheng[1,2], Rafal Klajn[1*]**

[1] Institute of Science and Technology Austria, Am Campus 1, 3400 Klosterneuburg, Austria

[2] College of Chemistry, Chemical Engineering and Materials Science, Soochow University, Suzhou, Jiangsu 215123, China

* Corresponding author; E-mail: carlo.rigoni@ist.ac.at, rafal.klajn@ist.ac.at

## Content:

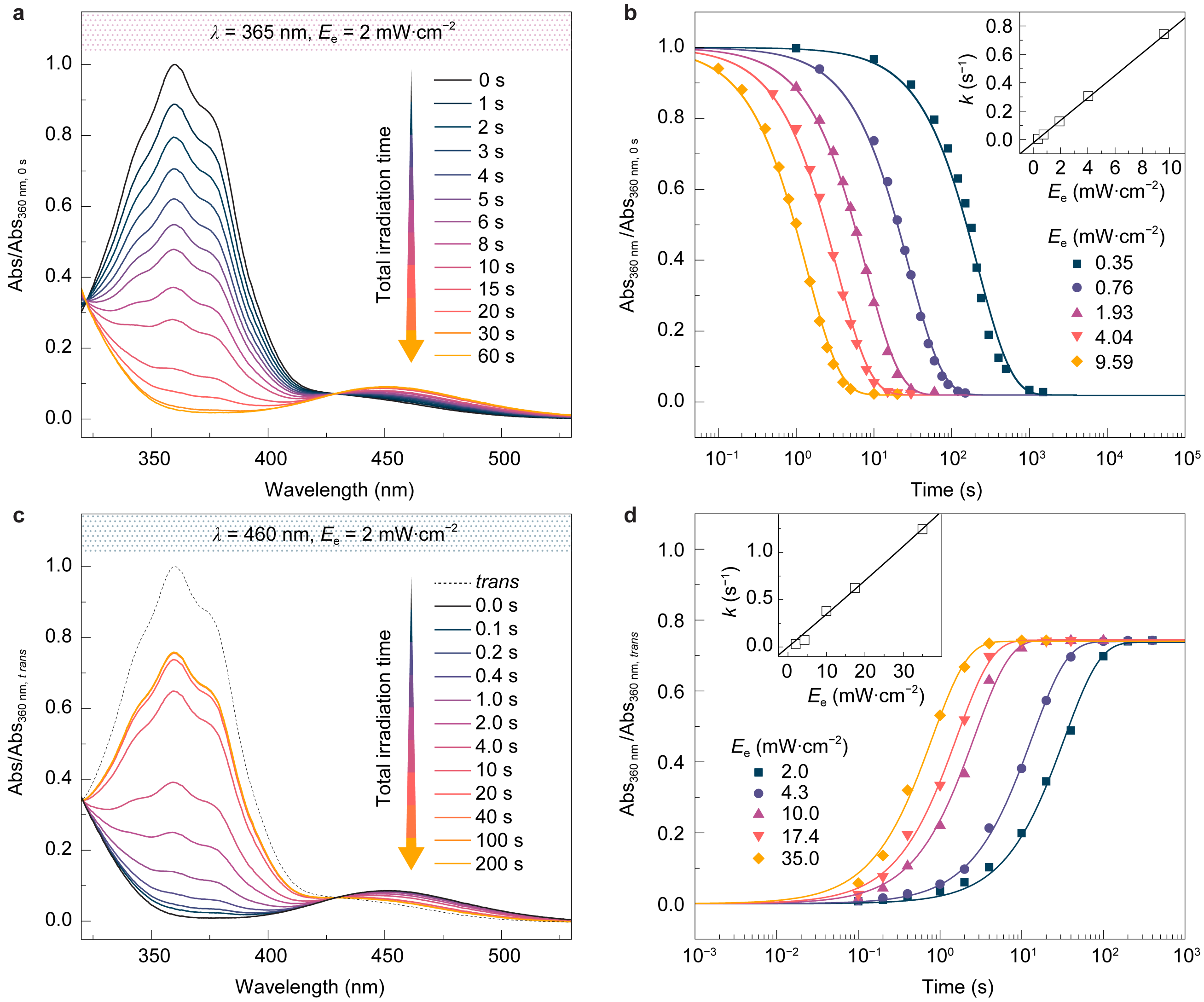

**Supplementary Fig. 1 | Analysis of the *trans*→*cis* and *cis*→*trans* photoisomerization of L-C8Azo by UV-vis absorption spectroscopy. a.** A representative series of UV-vis normalized absorption spectra for the *trans*→*cis* isomerization of L-C8Azo (15 μM in toluene) induced by irradiation with λ = 365 nm (irradiance, $E_e$ = 2 mW·cm$^{-2}$) **b.** Decrease of absorbance at 360 nm (normalized) due to *trans*→*cis* isomerization for different irradiances and irradiation times. The photoisomerization rate is linearly proportional to the irradiation intensity. **c.** A representative series of normalized UV-vis spectra for the *cis*→*trans* isomerization of L-C8Azo (15 μM in toluene) induced by irradiation with λ = 460 nm ($E_e$ = 2 mW·cm$^{-2}$). Spectra acquisition started immediately after establishing a *cis*-rich photostationary state (PSS) by prolonged irradiation with λ = 365 nm. **d.** Recovery of absorbance at 360 nm (normalized) due to *cis*→*trans* isomerization for different irradiances and irradiation times. The photoisomerization rate is linearly proportional to the irradiation intensity.

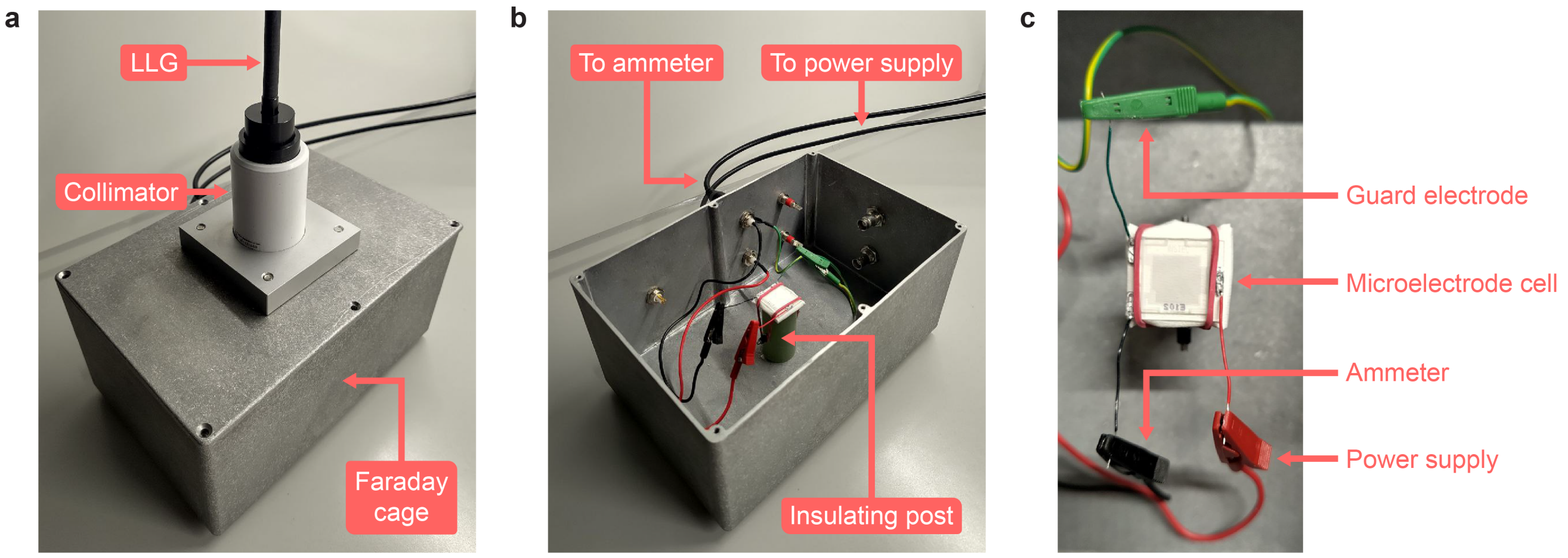

**Supplementary Fig. 2 | Photographic view of the experimental setup. a.** Photograph of the closed Faraday cage. The top lid of the cage has a see-through hole to accommodate the collimator and enable photoirradiation of the sample. The collimator is connected to the LED source through a liquid light guide (LLG). **b.** Photograph of the open Faraday cage. The microelectrode cell is mounted on top of an insulating post and connected to the ammeter and power supply. **c.** Close-up photograph of the microelectrode cell and the electrical wiring.

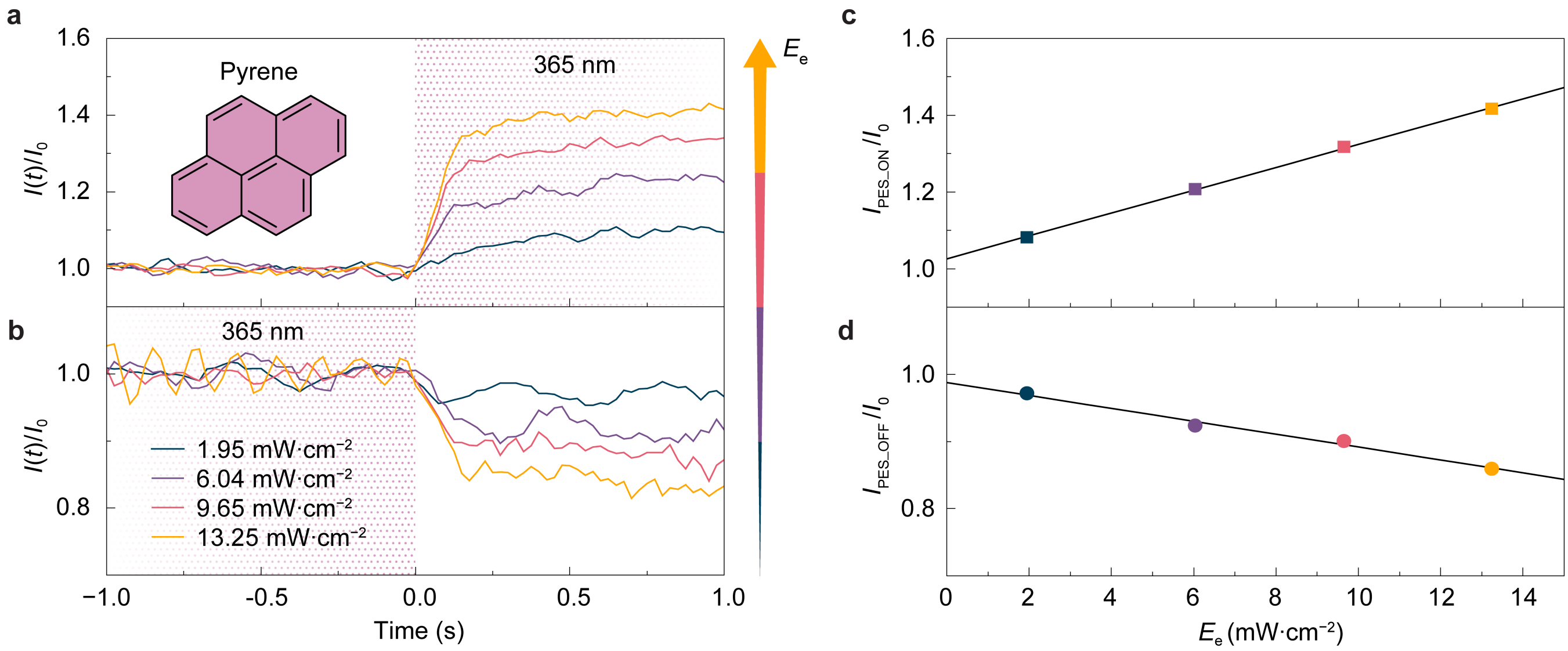

**Supplementary Fig. 3 | Effect of photoexcitation on the current for a 15 mM solution of pyrene in toluene. a.** Relative increase in the current when the light irradiation (at different irradiances) is turned on. **b.** Relative decrease in the current when the light irradiation (at different irradiances) is turned off. **c.** Initial jump in the current due to population of the photoexcited state (PES), extracted from the data in **a**. **d.** Sudden decrease of the current after switching off the irradiation, corresponding to the disappearance of the PESs, extracted from the data in **b**. Note: for each light intensity, the data in **a** and **b** correspond to the same experiment in which the voltage was first set to 2 V, then, after 300 s, the LED was turned on (data shown in **a**) for 300 s, then turned off (data shown in **b**) during a continuous current measurement. These data support the observation that PESs of non-photoisomerizable chromophores act as charge carriers in solution, leading to an increase in current (proportional to the light intensity).

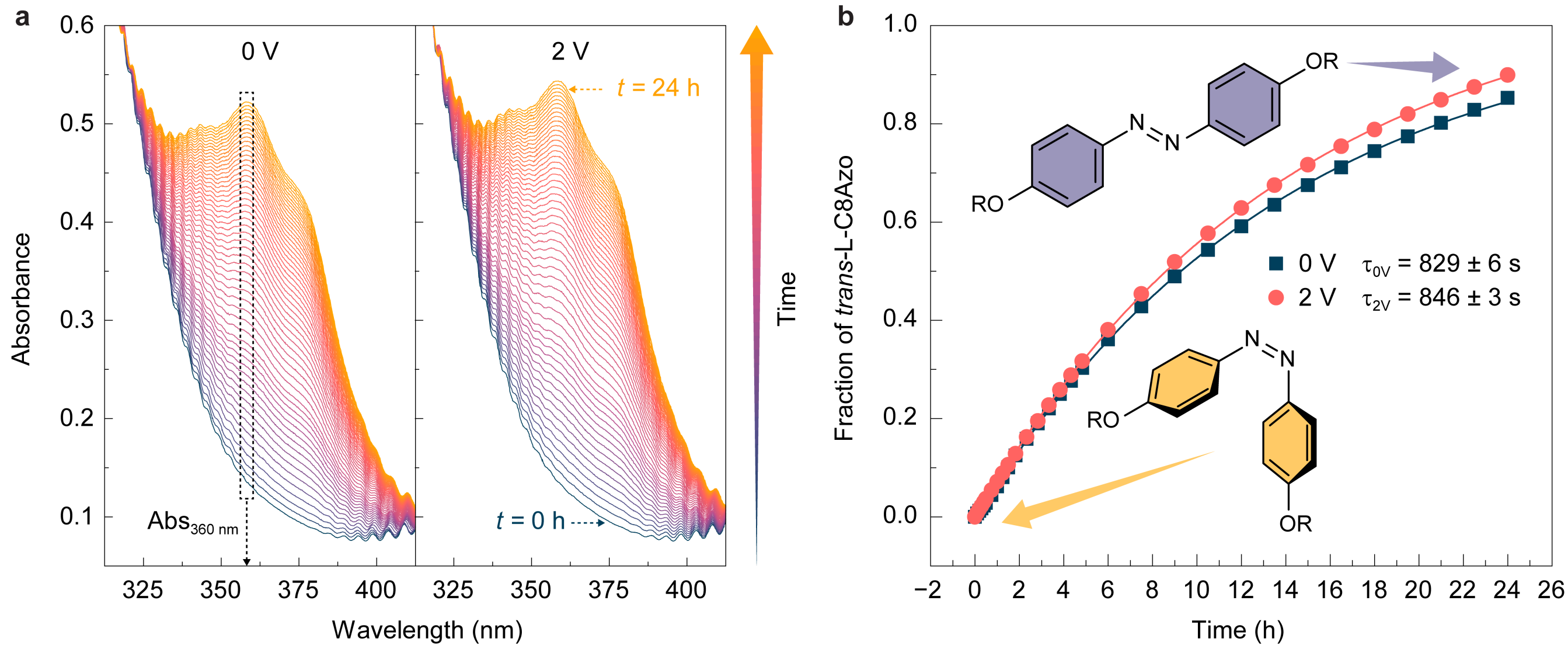

**Supplementary Fig. 4 | Effects of applying voltage on the *cis*→*trans* back-isomerization of L-C8Azo. a.** Time series of UV-vis absorption spectra accompanying the thermal back-isomerization of L-C8Azo in the absence of applied voltage (left) and under 2V (right). The interference pattern observable in the measurement is an artefact of performing these measurements directly in microelectrode cells. Spectra acquisition started immediately after establishing a *cis*-rich PSS by prolonged irradiation with $\lambda$ = 365 nm. **b.** Molar fraction of the *trans* isomer in the measurement cell over time, estimated from a calibration curve prepared by correlating the absorbance at 360 nm for different *trans*/*cis* mixtures with mixture composition (elucidated by NMR spectroscopy; see Supplementary Note 2).

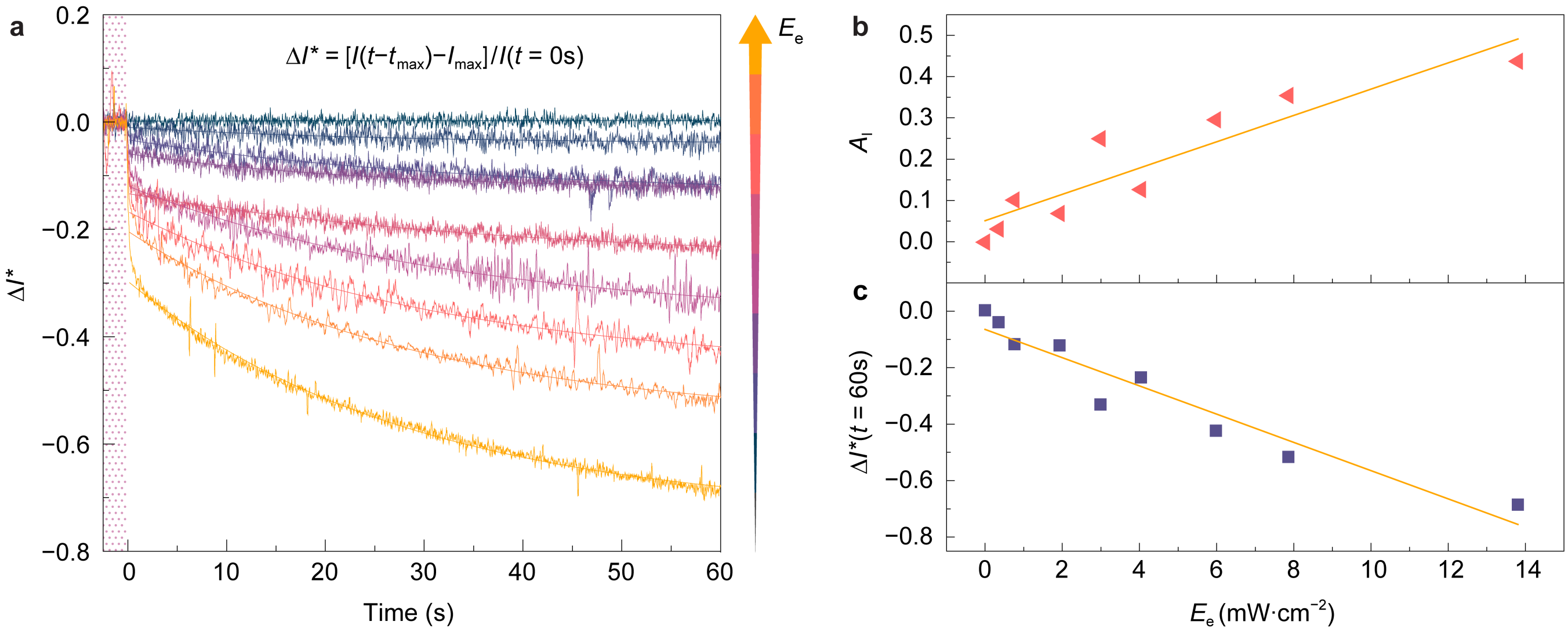

**Supplementary Fig. 5 | Effect of irradiation on the current decrease after turning off the LED. a.** Change in current as a function of time, measured after irradiating the sample with $\lambda$ = 365 nm at different irradiances. The data are normalized to remove the dependence on the absolute value of the current before starting the irradiation, enabling a direct comparison of the drop in current after irradiation has stopped. The datasets have been interpolated with a first-order exponential to correctly assess the initial step-drop due to vanishing contribution of molecules in the (PES) (see also Fig. 2e in the main text). **b.** Amplitude ($A_{\mathrm{I}}$) of the exponential decrease evaluated as $\Delta I^* = \Delta I_0^* + A_{\mathrm{I}} \cdot \exp[-t/t^*]$. The values of $t^*$ are similar for all the curves. **c.** Change in normalized current at $t$ = 60 s after the irradiation is turned off as a function of irradiation intensity. We hypothesize that both effects quantified in **b** and **c** are related to the decrease in conductivity associated with the decrease in temperature of the sample after irradiation is turned off. Upon irradiation, the sample and the surrounding glass reach different temperatures, due to photothermal effects on the sample and thermal equilibration with the environment. As soon as the irradiation is turned off, the fluid equilibrates back to room temperature, leading to a decrease in current.

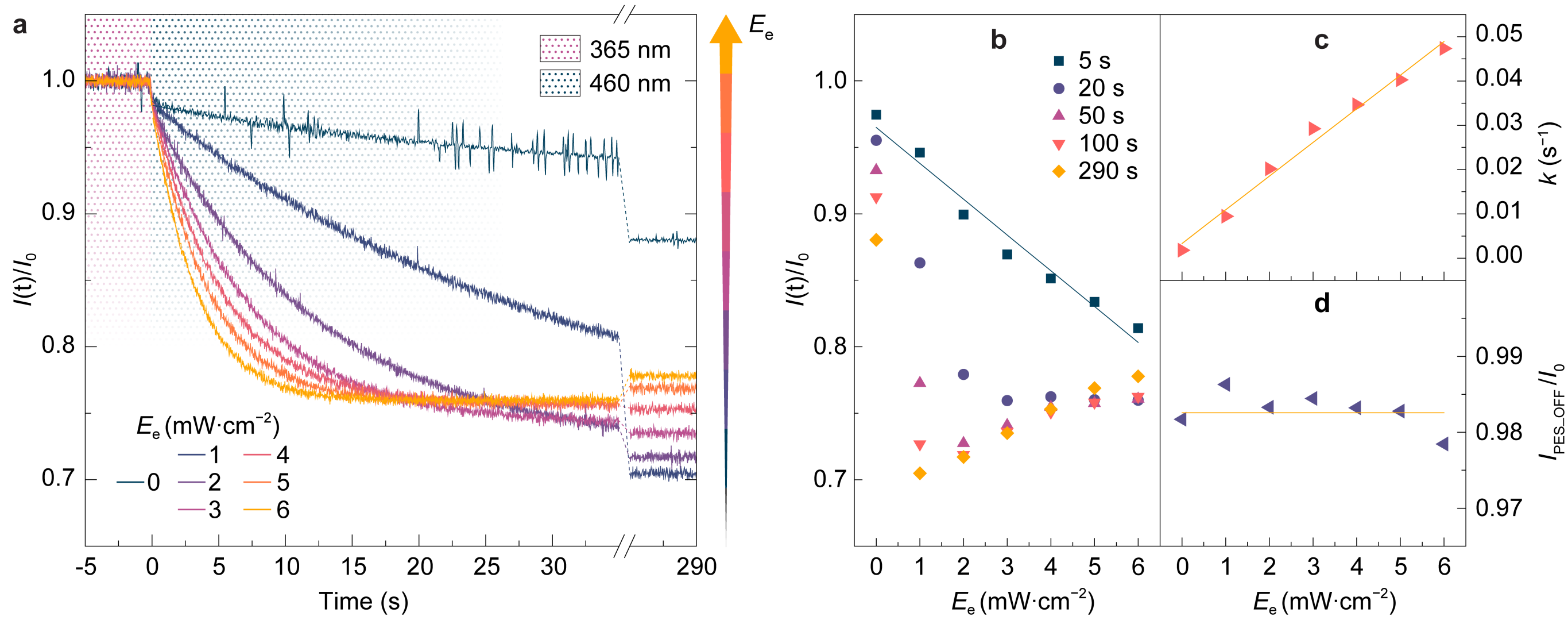

**Supplementary Fig. 6 | Effects of varying the irradiation intensity of the 460 nm LED. a.** Normalized current profiles obtained by irradiating a solution of L-C8Azo pre-exposed to UV light with $\lambda$ = 460 nm. After a step decrease (due to the disappearance of the L-C8Azo molecules in the UV-activated PES), the current first decreases linearly across all irradiances. Then, the current steadily decreases over time at lower irradiation intensities, while, after an initial increase, it slowly increases at higher irradiances. This final increase is probably due to photothermal effects from light absorption. **b.** Selected datapoints of the normalized current at different points in time and at different light intensities. In the short timescale ($t$ ~ 5 s), the current decrease is proportional to the light intensity. **c.** The reaction rate, measured as the slope close to the origin, scales linearly with irradiance. **d.** The current drop related to the inactivation of the UV-driven PES is independent of the $\lambda$ = 460 nm light applied afterward.

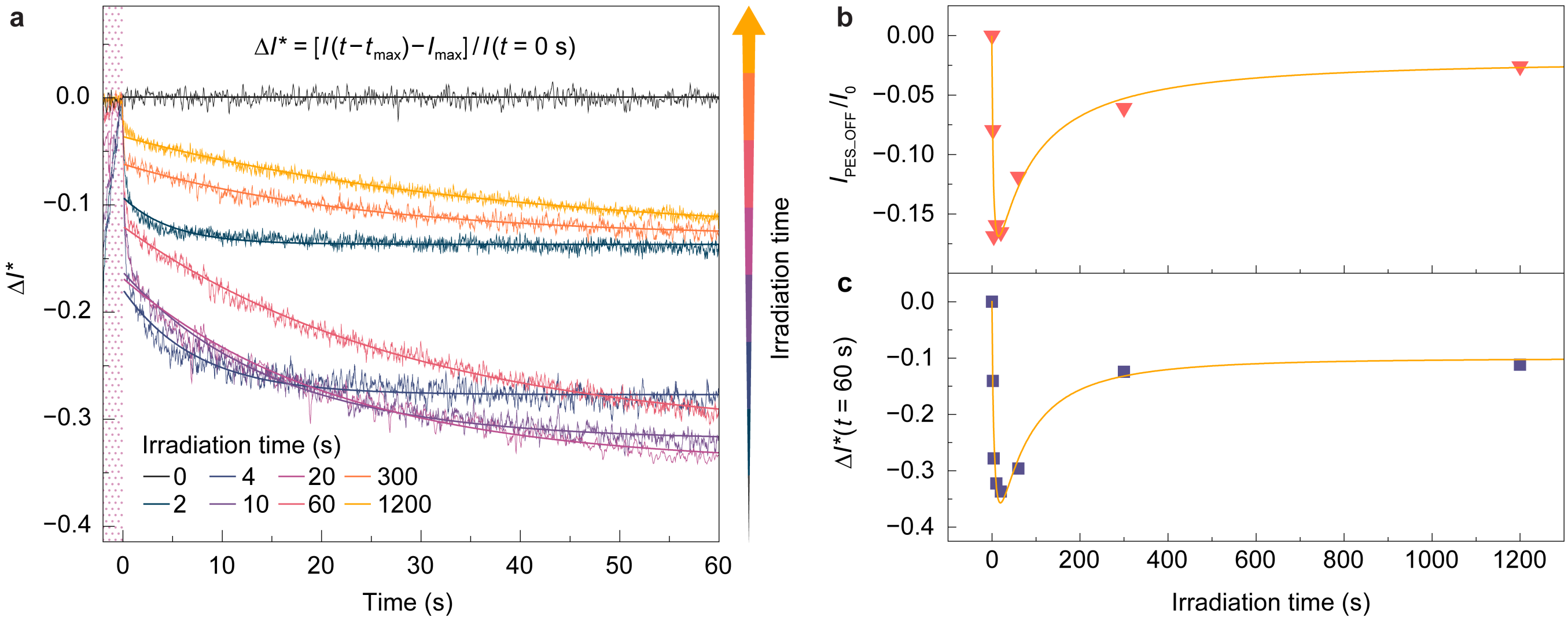

**Supplementary Fig. 7 | Effects of changing the irradiation time on the current decrease after turning off the LED. a.** Change in current as a function of time, measured after irradiating the sample for different irradiation periods. The data have been normalized to remove dependence on the absolute value of the current before irradiation, enabling a direct comparison of the post-irradiation current drop. The datasets have been interpolated with first-order exponentials to correctly assess the initial step-drop due to the vanishing of the contribution of the molecules in the PES to the current. **b.** Current drop associated with the disappearance of the PESs as a function of the irradiation time. **c.** Value of the renormalized current at $t$ = 60 s after the irradiation is turned off as a function of the irradiation time.

The behavior of the data in panels **b** and **c** can be rationalized as follows. In **b**, the drop in current due to the vanishing contribution of the molecules in the PES depends on how many molecules in the *trans* state are available for photoexcitation with $\lambda$ = 365 nm. The longer the irradiation time, the fewer *trans* molecules remain in the mixture due to ongoing photoisomerization (until a photostationary state is reached). As a result, the drop becomes smaller as the irradiation time increases. This decrease can be seen for the data for 60 s, 300 s, and 1200 s of irradiation time. For the data for 4 s, 10 s, and 20 s, this effect is negligible. If the contribution of the photoisomerization were the only factor, however, the data at 2 s should have the same value as the ones up to 20 s, which is not the case. We hypothesize that the inconsistency for the value at 2 s is due to the fact that with only 2 s of irradiation, the temperature of the fluid is much lower than at higher irradiation times. The impact of temperature is much more evident in panel **c**. Here, the exponential decrease is mostly driven by decreases in temperature. For the 2-s irradiation trace, the decrease is much smaller because the irradiation time is not long enough to heat up the fluid. For 300 s and 1200 s, the decrease is smaller, because such long irradiation times are enough to start changing the temperature of the glass surrounding the fluid, leading to smaller thermal equilibration effects after the LED is switched off.

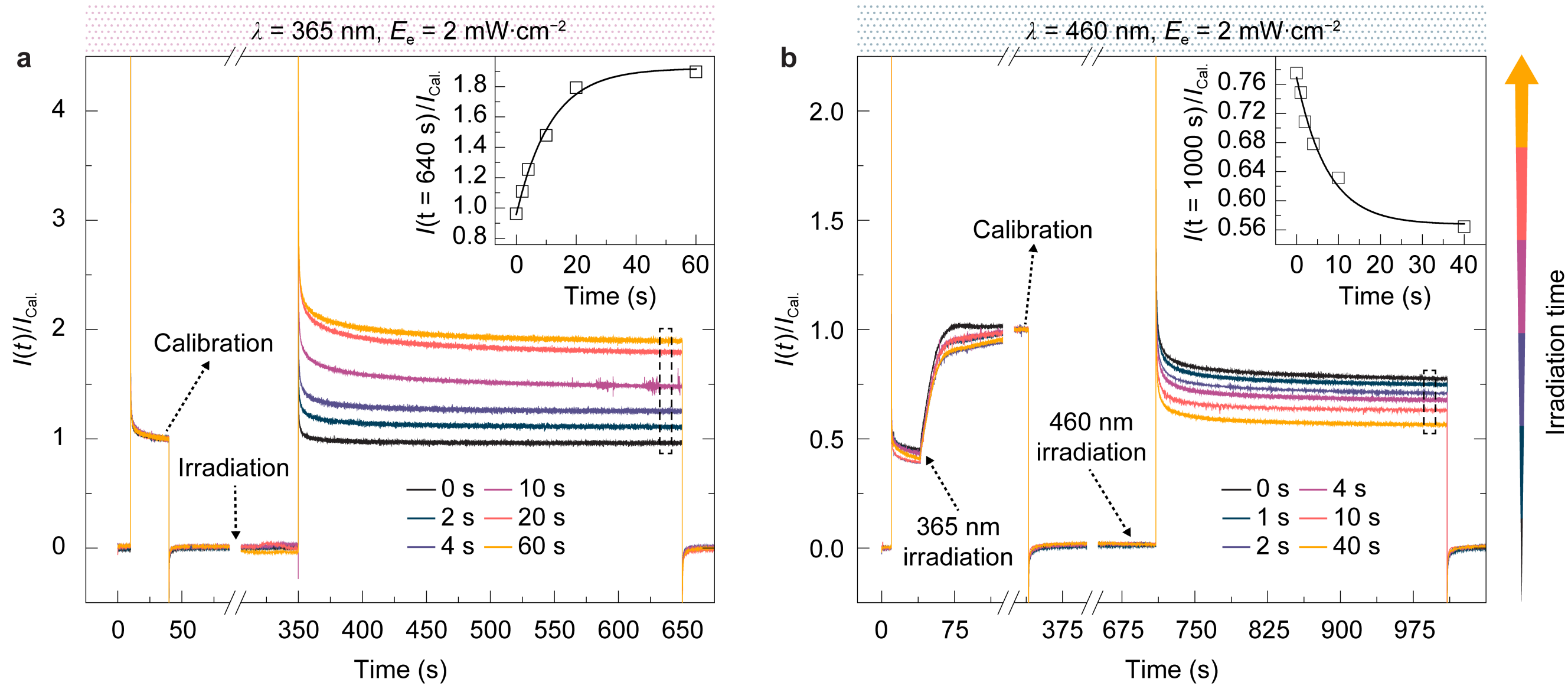

**Supplementary Fig. 8 | Effects of changing the irradiation time if the irradiation occurs when the voltage is not applied.** **a.** Change in current as a function of time, measured to assess the effect of the irradiation length on conductivity (*trans→cis* isomerization). First, a voltage (2 V) is applied to evaluate the steady-state current in the *trans* state in each case (calibration current, $I_{Cal.}$). Then, the voltage is turned off and the cell is irradiated at $\lambda$ = 365 nm and $E_e$ = 2 mW·cm$^{-2}$ for different durations. After irradiation, the voltage is kept off for a couple of minutes, then turned back on to the same value as before. The steady-state current during the second voltage step is then measured and plotted (inset). The current increases exponentially with the irradiation time, in accordance with the *trans→cis* isomerization dynamics. **b.** Current as a function of time, measured to assess the effect of the irradiation length on the conductivity (*cis→trans* isomerization). First, a voltage (2 V) is applied, and after a brief equilibration time, the cell is irradiated with $\lambda$ = 365 nm at $E_e$ = 2 mW·cm$^{-2}$ for 300 s to evaluate the steady-state current in the sample when it is saturated with the *cis* isomer (calibration). Then, the voltage is turned off, and the cell is irradiated with $\lambda$ = 460 nm at $E_e$ = 2 mW·cm$^{-2}$ for different durations. After irradiation is turned off, the voltage is kept off for a couple of minutes, then turned back on at the same value as before. The steady-state current during the second voltage step is then measured and plotted (inset). The current decreases exponentially with the irradiation time, in accordance with the *cis→trans* isomerization dynamics.

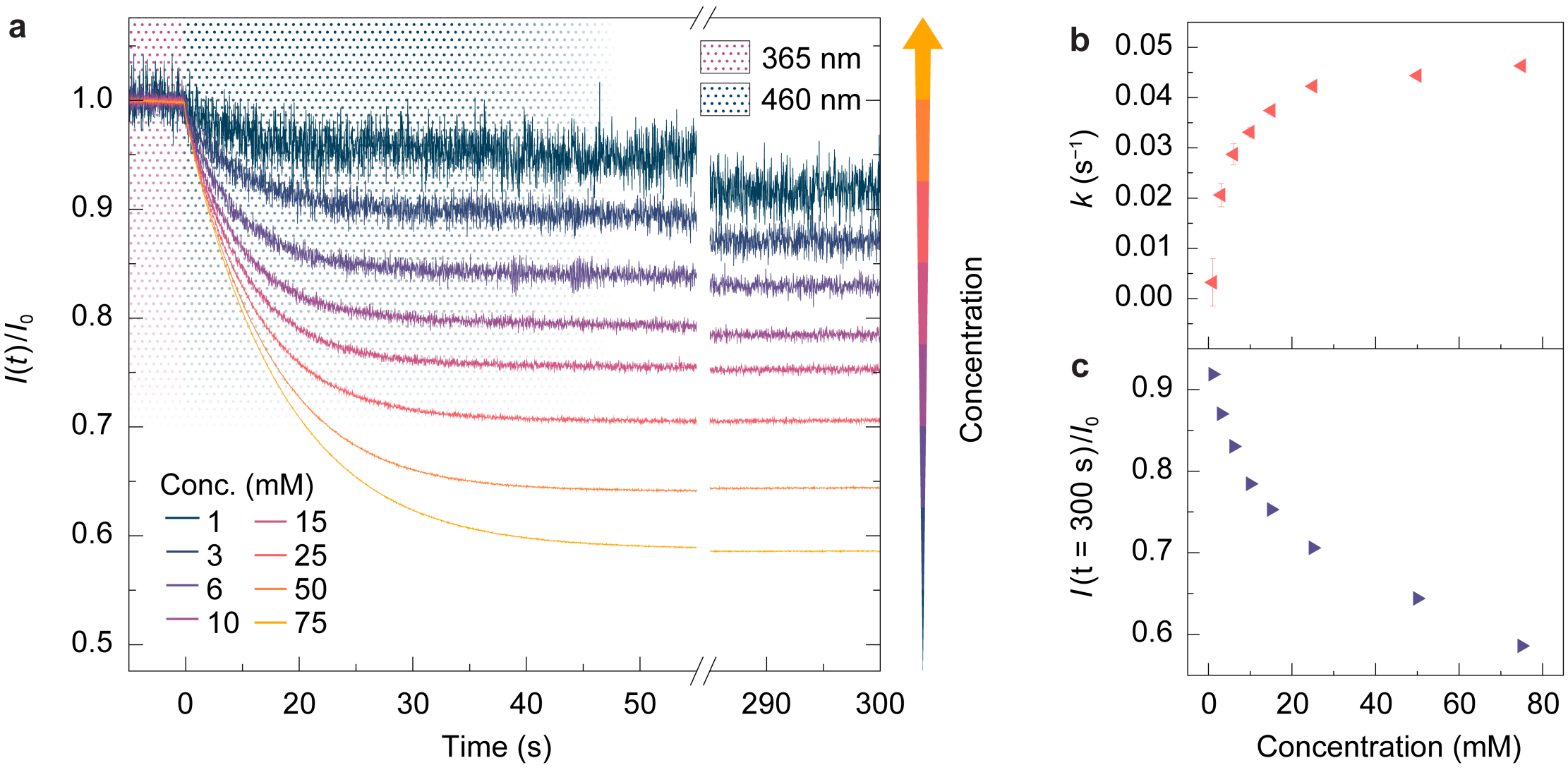

**Supplementary Fig. 9 | Effects of varying concentration for samples subjected to λ = 460 nm. a.** Current profiles obtained by irradiating L-C8Azo pre-exposed to UV light with λ = 460 nm. **b.** Nonlinear increase of the photoisomerization rate with concentration. **c.** Increasingly prominent relative current drop at high concentrations.

The increase in reaction rate with concentration is due to the temperature increase in the fluid at a higher concentration, due to stronger light absorption. The apparent saturation of the photoreaction rate (**b**) is probably related to the decreased penetration depth in the sample due to high concentration.

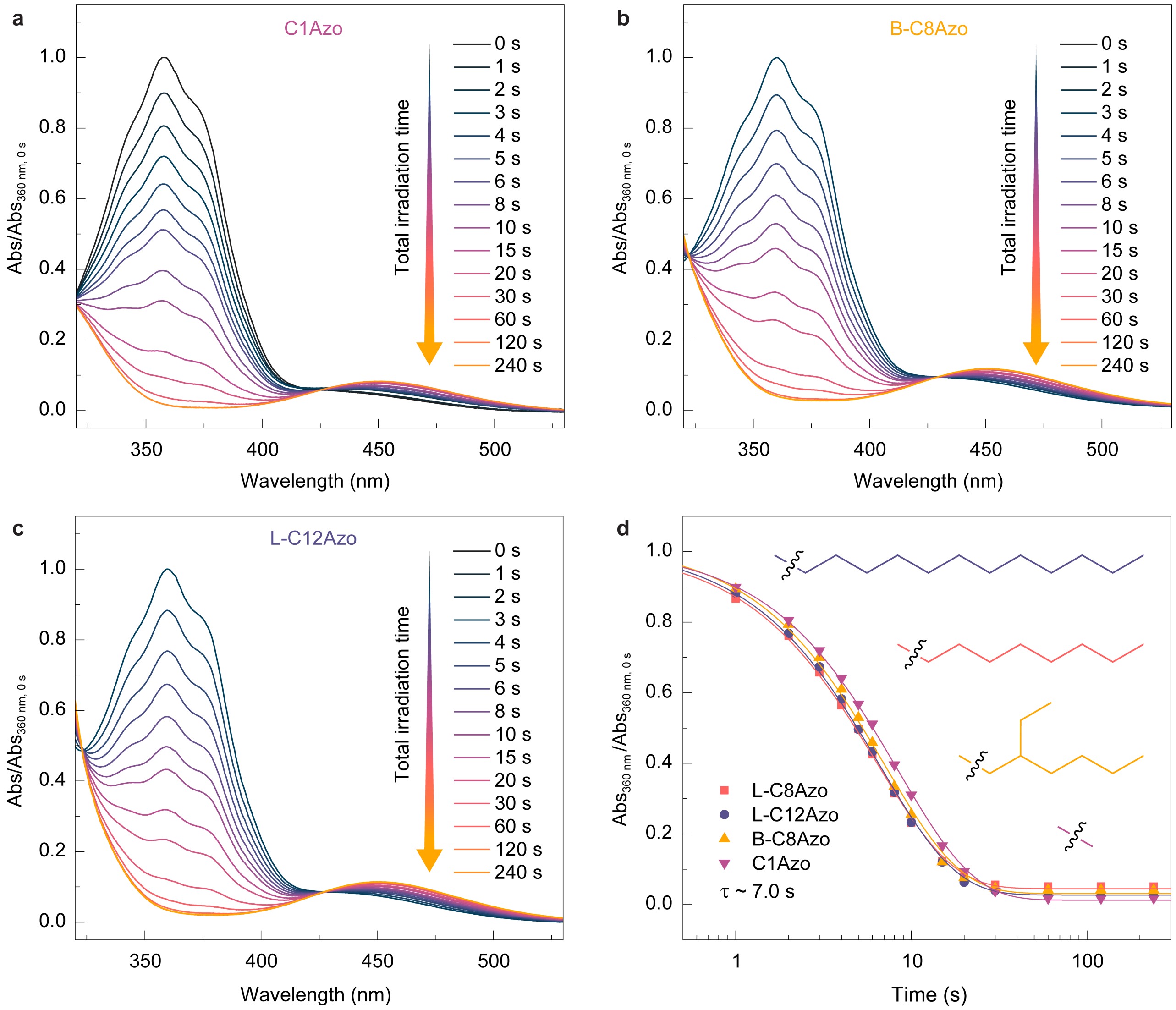

**Fig. S9 | Analysis of the *trans*→*cis* photoisomerization of the differently substituted azobenzenes. a–c.** Changes in UV-vis absorption spectra upon *trans*→*cis* isomerization of (**a**) C1Azo, (**b**) B-C8Azo, and (**c**) L-C12Azo. **d.** Decrease in absorbance at 360 nm for L-C8Azo, C1Azo, B-C8Azo, and L-C12Azo during the *trans*→*cis* isomerization. Similar characteristic timescales of transition ($\tau$) indicate negligible effect of the *para*-substituents on photoswitching properties. All photoisomerization experiments were performed on 15 μM toluene solutions in cuvettes with a 1-cm path length; irradiation conditions: $\lambda$ = 365 nm, $E_e$ = 2 mW·cm$^{-2}$.

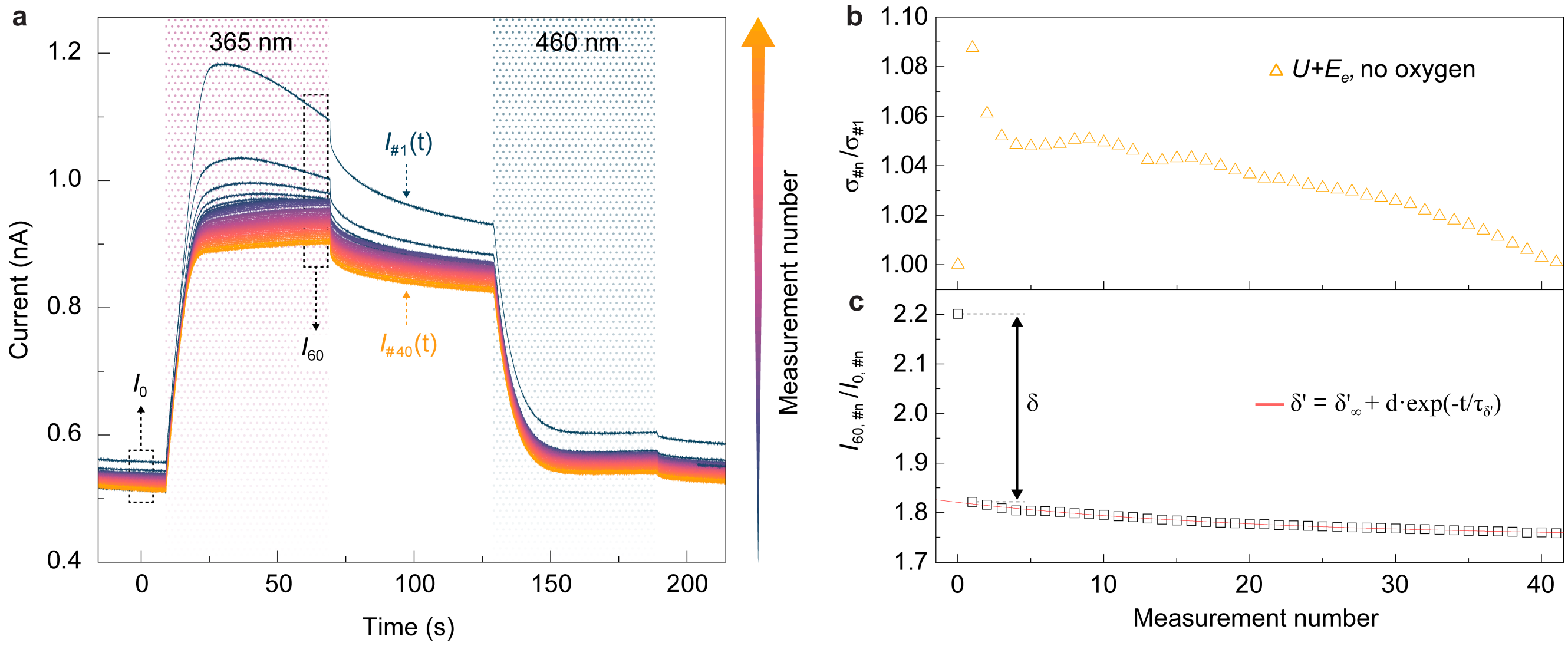

**Supplementary Fig. 11 | Evolution of the system's behavior during repeated photoswitching in absence of oxygen. a.** Graph of the current data of 41 consecutive current measurements under repeated cycles of UV followed by blue light irradiation with a 15-min pause between the cycles. **b**. Evolution of conductivity at measured for each repetition immediately before the light irradiation is turned on (*t* = 0 s) for consecutive runs under irradiation and voltage and in absence of oxygen. **c**. Relative increase of the photostationary state current ($I_{PSS}$) for consecutive measurements.

Even if the sample is prepared under argon atmosphere, small traces of oxygen are still present in solution. As the oxygen is the limiting factor, the first cycle consumes most of the trace oxygen present, especially in the first 60 s where the current visibly drops even during the 365 nm irradiation stage. The consumption of oxygen during the first cycle causes the second cycle to have an overall lower current than the first, masking even the effect of the loss of switching efficiency due to the incomplete back-isomerization of L-C8Azo. This, however, does not modify the overall switching performance within each single cycle for which the current still increase of ~2-fold factor upon irradiation with 365 nm.

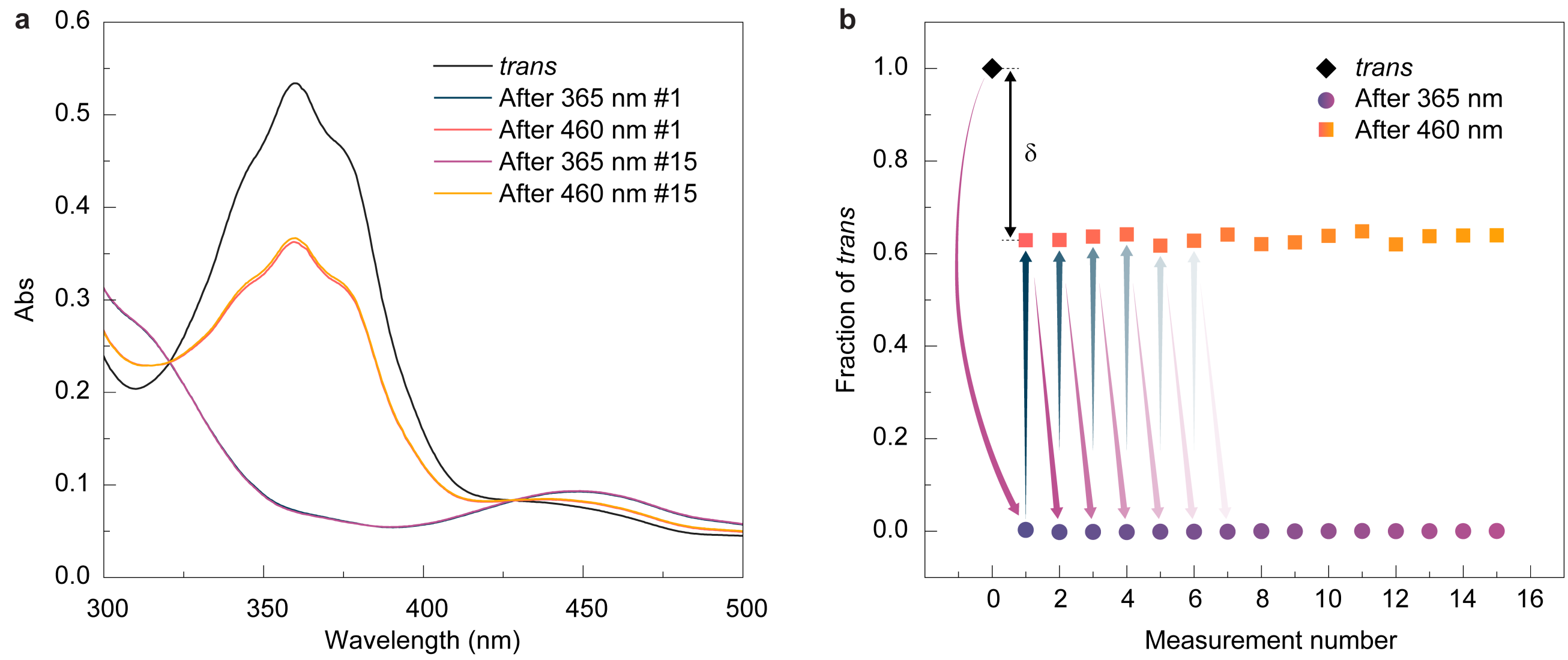

**Supplementary Fig. 12 | Effect of repeated photoswitching of L-C8Azo using 365 nm and 460 nm irradiation on the UV-vis absorption spectra.** The measurement protocol was defined by a series of irradiation and UV-vis measurement steps. Once the initial UV-vis spectrum was recorded (the *trans* isomer), the sample was irradiated with alternating cycles of λ = 365 nm (10 s at 10 mW·cm$^{-2}$) and λ = 460 nm (10 s at 20 mW·cm$^{-2}$). After a given number of cycles, the spectrum was re-recorded. **a.** UV-vis absorption spectra obtained during and after at the first (#1) and fifteenth (#15) irradiation cycle. **b.** Molar fraction of molecules in the *trans* form as a function of irradiation cycle. The fractional content of *trans*-L-C8azo was determined by calibrating the absorbance values using NMR spectroscopy (see Supplementary Note 2). After the incomplete recovery of the *trans* isomer after the first switching cycle (δ) due to the inefficient blue-light-driven *cis*→*trans* back-isomerization, the fractional content of *trans*-L-C8Azo oscillates between ~0 and ~60 mol% without noticeable fatigue.

## Supplementary Note 1: Physical modeling of photoconductivity

In this Note, we build a simple model for photoconductivity changes associated with the isomerization of L-C8Azo. We begin by arguing that *cis*-L-C8Azo is more efficient at exchanging charges with the electrodes than *trans*-L-C8Azo and show that this simple hypothesis is enough to explain the exponential profile of the current increase upon light irradiation. Furthermore, we describe how the charge exchange and electrophoresis govern the phenomenology of the electric current readings in the system.

### 1.1 Charge exchange at the electrode: *trans* vs *cis*

The charge-exchange phenomenon occurs through tunneling of electrons from the metal surface to molecules in the fluid[1]. As a consequence, the efficiency of charge exchange depends on i) the distance of the molecule from the electrode, and ii) the difference in the energy of the electronic states in the metal and the molecule.

In contrast to molecules in the *trans* state, in which only transient short-lived electrical dipoles are present, those in the *cis* state (for symmetrically *para*-substituted azobenzenes, like the ones used in this study), have measurable permanent electrical dipoles ($p \approx 4.4$ D for the *cis* state vs ~0 D for *trans*[2]). This permanent dipole makes the molecule prone to reorient along the electric field lines and mediates attracting charge–dipole forces between the molecule and inhomogeneities on the surface that generate local gradients of electric field close to the electrode. These factors increase the likelihood that a molecule in the *cis* state will come close to the electrode and, consequently, increase its rate of charge exchange. Furthermore, when the electric field is turned on, the impurities present in solution migrate toward the electrodes, and, even at low concentrations, induce the development of a more polar environment there rather than in the bulk. Being slightly polar, the molecules in the *cis* state have easier access to the electrodes than the molecules in the *trans* state, which are instead repelled. Finally, electrons in the highest occupied molecular orbitals of *cis*-azobenzenes are higher in energy than those of *trans*-azobenzene, making it even easier to become charged through tunneling, as the energy needed for them to accept or lose an electron is smaller than the one needed from the *trans* molecules.

Following this argument, charge exchange naturally occurs at a higher rate in solutions that have higher concentrations of molecules in the *cis* state. Therefore, the conductivity of the solutions of L-C8Azo in toluene depends not only on the concentration of photoresponsive molecules in the mixture but also on the ratio between molecules in the *trans* and in the *cis* states.

## 1.2 Base assumptions and experimental conditions

To model the conductivity as a function of the *trans→cis* isomeric distribution in the mixture, let us first make assumptions based on the experimental conditions in our study.

*a. First-order kinetics*

First, let us suppose that the photoisomerization reaction between *trans*- and *cis*-L-C8Azo follows first-order kinetics. This assumption is true if either the compound's concentration is low or only a thin layer of it is irradiated, allowing the entire solution to be irradiated with the same light intensity at the same time. This condition is satisfied in all our experiments with the exception of the ones done at the higher concentrations of L-C8azo (>15 mM; see Fig. 4 in the manuscript and later discussion in this note).

*b. Concentration–conductivity linearity*

Second, let us assume that the increase in conductivity of mixtures of L-C8Azo in toluene at low concentrations is linear with the increase in concentration of the *cis* molecules. Despite an evident nonlinearity in the entire range of concentrations studied, at low concentration, the increase in conductivity is fairly linear with the concentration of L-C8Azo (see Fig. 4). Since at low concentration, we can assume that intermolecular interactions are negligible, assuming linearity between the number of molecules in the *cis* state and the conductivity is reasonable.

*c. Negligible capacitance change*

Third, let us assume that the change in dipole moment induced by the L-C8Azo *trans→cis* isomerization does not affect the overall capacitance of the fluid. To assess this assumption, we can start by calculating the impact of the *trans→cis* isomerization on the overall average electrical dipole of the mixtures. For a 15 mM solution, the ratio between L-C8Azo and toluene molecules in solution is ~1.6:1000. The electric dipole moment of toluene and *cis*-L-C8Azo is ~0.3 D and ~4.4 D, respectively. As such, if all L-C8Azo molecules in solution undergo the *trans→cis* isomerization, the final average dipole moment would increase by a theoretical maximum of 1.6% with respect to the initial dipole. Furthermore, despite the lack of a simple linear relationship between the average dipole moment of a mixture of molecules and its dielectric constant, the two are closely related. It is therefore reasonable to assume that, in the absence of large collective interactions between the molecules as in the dilute concentration case, a negligible change in the overall average dipole moment of the mixture would not affect the dielectric constant in a major way. Finally, as the capacitance of the fluid is linearly dependent on the dielectric constant as $C = \varepsilon A/d$, we can also assume that the *trans→cis* isomerization does not affect the overall capacitance of the system. To verify this last assumption experimentally, we measured the current response to a triangular voltage sweep for a 15 mM solution of L-C8Azo in toluene (Supplementary Fig. 13), showing that the contribution of the capacitance, estimated from the difference in the current between the increasing and decreasing branches of the voltage sweep, is similar in the *trans* and *cis* states.

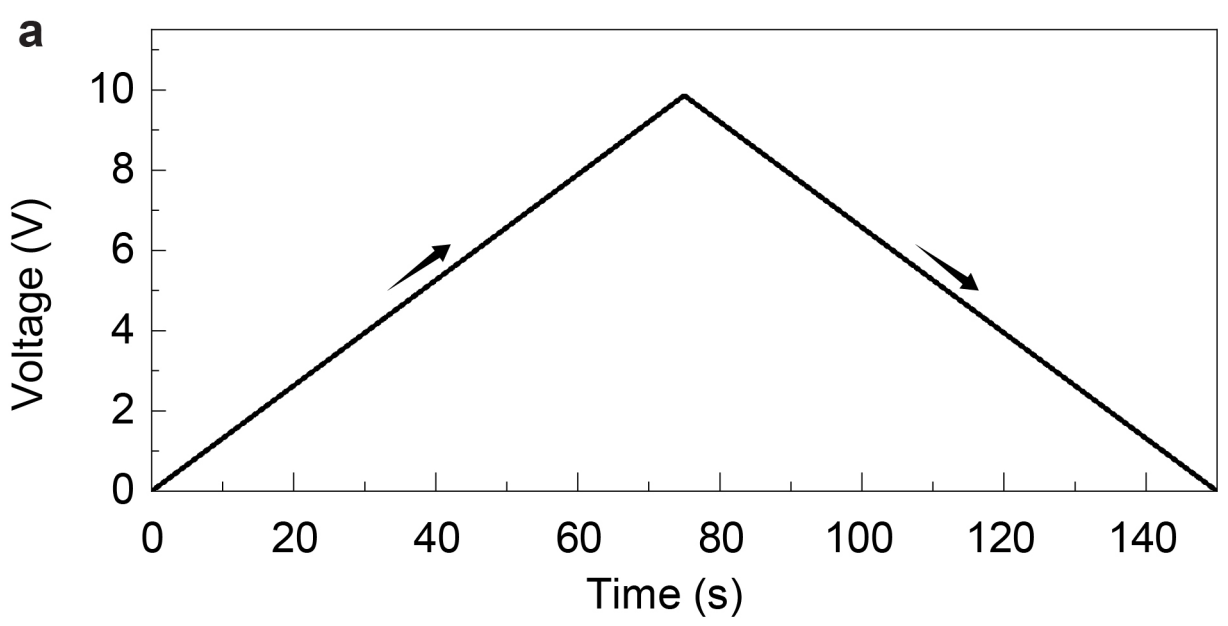

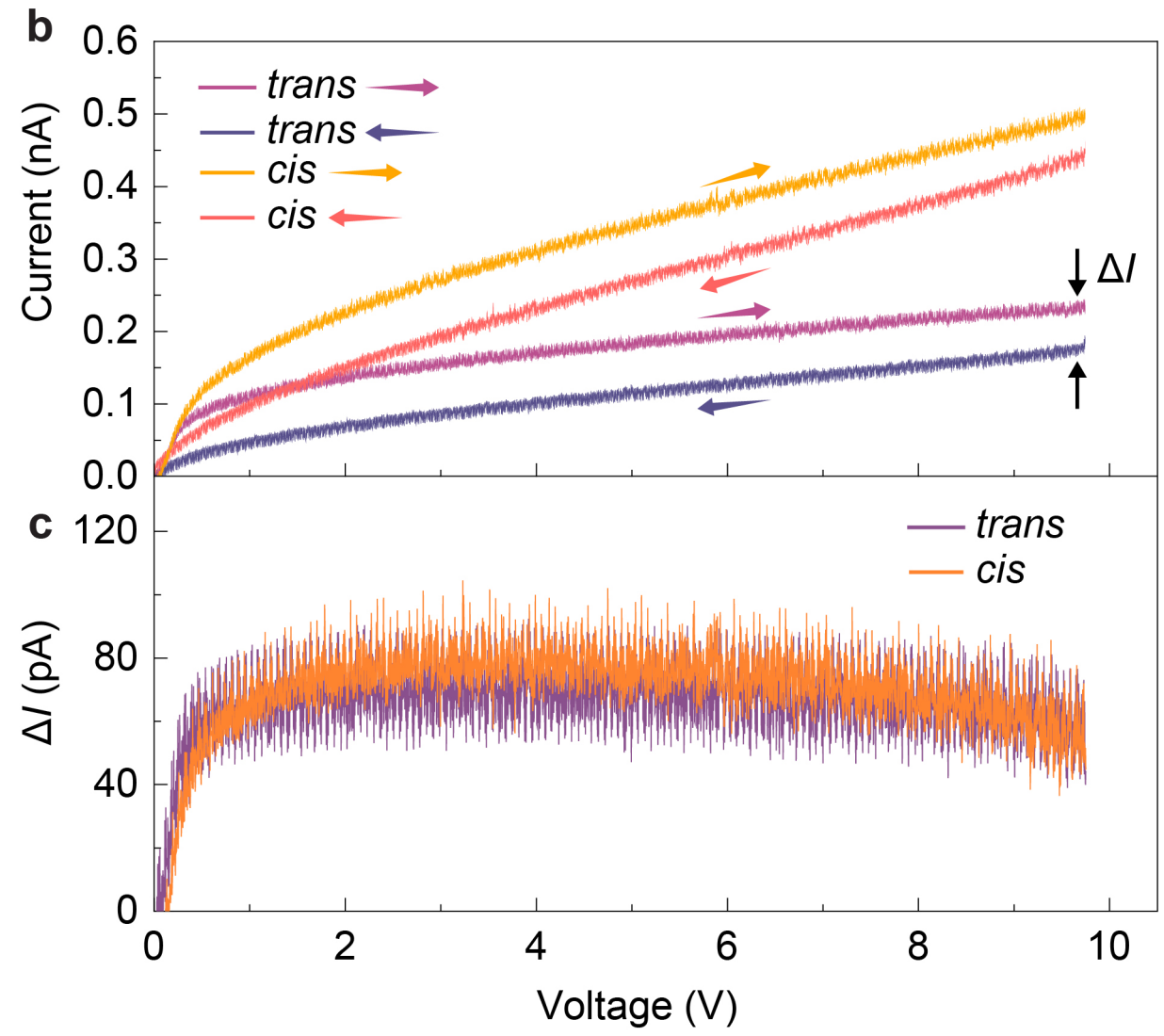

**Supplementary Fig. 13 | Effect of photoirradiation on capacitive effects. a.** Voltage increase over time used in the following measurements. **b.** Current obtained by increasing and decreasing the voltage. **c.** Delta in the current at each voltage applied between the increasing and decreasing profiles of the curve. The delta is dependent from the capacitive component in the circuit.

## 1.3 From isomerization to conductivity

Let us consider the following initial concentrations of the *trans*- (species A) and *cis*- L-C8Azo (species B) isomers:

$$t = 0, \quad [A] = [A]_0, \quad [B] = 0; \tag{1}$$

such that the molecules are all in the *trans* state at $t = 0$. Following assumption *a,* the concentration of species A and B during *trans→cis* isomerization over time can be expressed as:

$$[A] = [A]_0\, e^{-kt}, \tag{2}$$

$$[B] = [A_0](1 - e^{-kt}), \tag{3}$$

where $k$ is the reaction rate. Following assumption *b,*

$$\sigma_\mathrm{A} = \alpha[A], \quad \sigma_B = \beta[B]; \tag{4}$$

$$\sigma = \sigma_\mathrm{A} + \sigma_\mathrm{B}. \tag{5}$$

Combining eqs. (2) and (3) with (4) and (5) we obtain:

$$\sigma(t) = \beta[A_0] + [A_0](\alpha - \beta)e^{-kt}. \tag{6}$$

It is worth noting that:

$$\lim_{t\to 0} \sigma(t) = \alpha[A_0] = \sigma_\mathrm{A}, \tag{7}$$

$$\lim_{t\to +\infty} \sigma(t) = \beta[A_0] = \beta[B_\infty] = \sigma_\mathrm{B}. \tag{8}$$

Finally, following assumption *c*, the current $I \sim \sigma$ at constant voltage. Therefore, Equation 6 captures the exponential growth of the current obtained by irradiating light in our system. The equations above work symmetrically for the *cis→trans* back isomerization.

## 1.4 Limits of the model

To verify the limits of the simple model obtained in Equation 6, we have interpolated the current data corresponding to the beginning of the irradiation window with $\lambda$ = 365 nm at different concentrations (Supplementary Fig. 14). From the analysis, the model accurately describes the observed behavior up to a concentration of 15 mM of L-C8Azo in toluene. At higher concentrations, the plot of the residuals (Supplementary Fig. 14b) and their distribution (Supplementary Fig. 14c) reveal that the model no longer fully capture the photoconductive behavior. In particular, the higher the concentration the larger the discrepancy between the data and the model. As mentioned in the main text, the phenomenon is most probably due to the breaking of assumption $a$, meaning that the isomerization can no longer be described as a first order process due to its high absorbance. In short, based on the Beer–Lambert law: absorbance $A = \varepsilon c l$, where $\varepsilon$ (the molar extinction coefficient) for azobenzenes[3] is approx. 25 000 $M^{-1}\cdot cm^{-1}$, $c = 75$ mM, and $l = 10$ μm, therefore $A = 1.875$. Now, $A = \log(\Upsilon_0/\Upsilon)$, where $\Upsilon_0$ is the intensity of incident light and $\Upsilon$ is the intensity of transmitted light. For $A = 1.875$, the ratio $\Upsilon_0/\Upsilon = 10^{-1.875} = 0.013$. In other words, around 1.3% of the incident light passes through the sample. Most of the light is therefore absorbed by the top layer, and only a fraction filters through to the bottom layer.

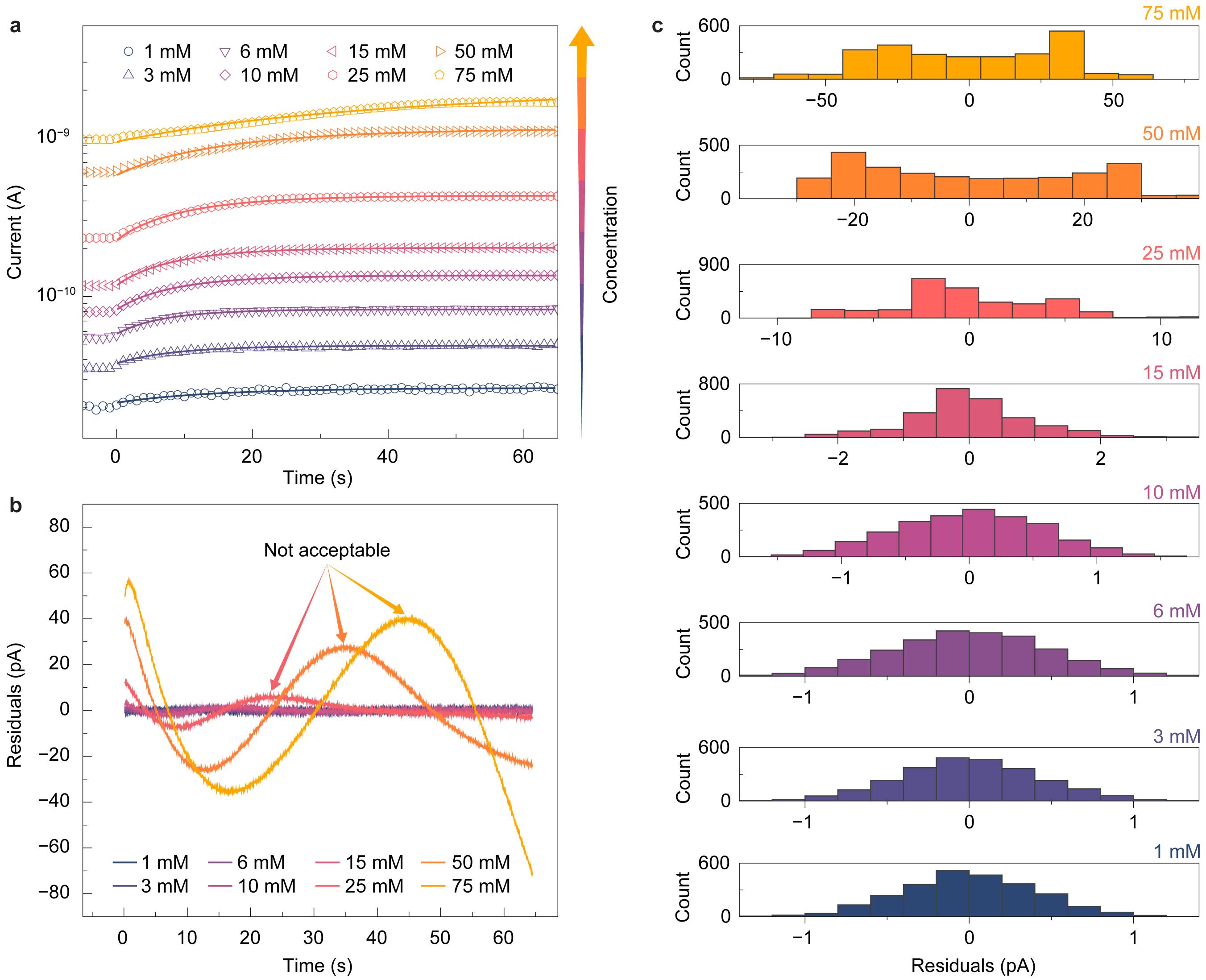

**Supplementary Fig. 14 | Testing the limits of the model relating photoisomerization to conductivity. a.** Current data at different concentrations interpolated with Equation 6. **b.** Plot of the residuals from the interpolation of the model at different concentrations. **c.** Distribution of the residuals from the interpolation of the model at different concentrations.

## 1.5 Current–voltage profiles

In this last part, we focus on the interpretation of the current–voltage behavior (Fig. 1d). From the collected experimental data, we can identify two distinct regimes: first, the current increases non-linearly at low voltages, and then it increases linearly. The behavior observed in these two regimes is dependent on two different mechanisms working in parallel: charge exchange at the electrodes via tunneling, and charge transport across the cell via electrophoresis. The details regarding the dependence of the current from the contribution of the charge exchange via tunneling will be carried out below adapting the theory developed in a recent work by Bevan K. H. *et al.* (Ref. 1).

*Tunneling probability*

Depending on the direction of electron transfer, tunneling at the electrode results in the molecular undergoing either reduction of oxidation (for the electron-rich L-C8Azo the latter reaction is more likely). The energy states of the excited reactant accessible for these reactions are described by the Gerischer–Hopfield distribution[1]:

$$D_{\mathrm{ox,red}}(\varepsilon) = \frac{1}{\sqrt{4\pi\lambda k_{\mathrm{B}}T}} \exp\left(-\frac{\left(\varepsilon - \varepsilon_{\mathrm{ox,red}}\right)^2}{4\lambda k_{\mathrm{B}}T}\right) \tag{9}$$

where $\lambda = (\varepsilon_{\mathrm{ox}} - \varepsilon_{\mathrm{red}})/2$, $\varepsilon_{\mathrm{ox,red}}$ are the average oxidation and reduction energies, respectively, $k_{\mathrm{B}}$ is the Boltzmann constant, and $T$ is the temperature.

At equilibrium, when the electrode's electrochemical potential $\mu_{\mathrm{eq}} = \varepsilon_{\mathrm{red}} + \lambda = \varepsilon_{\mathrm{ox}} - \lambda$, electron tunneling to oxidation or reduction of the molecule is equally unlikely, resulting in zero current[1]. As soon as a bias voltage is applied, $\mu_{eq}$ is shifted to $\mu = \mu_{eq} + U$, and the probability of oxidation of a reactant molecule becomes larger at the positive electrode and smaller at the negative electrode. The opposite is true for reduction. The forward ($k_{\mathrm{f}}$) and backward ($k_{\mathrm{b}}$) rates of the reactions occurring at the opposite electrodes can be expressed as[1]:

$$k_{\mathrm{f}} = \frac{4\pi|M^2|}{2\pi\hbar} D_{\mathrm{s}} \int f(\varepsilon) D_{\mathrm{ox}}(\varepsilon) d\varepsilon \tag{10}$$

$$k_{\mathrm{b}} = \frac{4\pi|M^2|}{2\pi\hbar} D_{\mathrm{s}} \int [1 - f(\varepsilon)] D_{\mathrm{red}}(\varepsilon) d\varepsilon, \tag{11}$$

where $M$ is the electronic coupling between single-particle states in the electrode and reactant, $D_s$ is the density of states in the electrode, $\hbar$ is the reduced Planck constant, and

$$f(\varepsilon) = 1/\left[1 + \exp\left(\frac{\varepsilon - \mu}{k_{\mathrm{B}}T}\right)\right] \tag{12}$$

is the Fermi function, where the dependence of the rates on the electrode's electrochemical potential $\mu$, and therefore on the applied voltage, becomes evident.

The oxidation/reduction probability continues to increase with the increase of the voltage, making $k_{\mathrm{f}} > k_{\mathrm{b}}$ until all the energy states for the oxidation/reduction processes become accessible ($\mu \gg \mu_{\mathrm{eq}} + 2\lambda$ or $\mu \ll$ $\mu_{\mathrm{eq}} - 2\lambda$). From this point onwards, all the L-C8Azo molecules that come into contact with the electrodes become oxidized or reduced with $k_{\mathrm{b}} \rightarrow 0$ and:

$$k_{\mathrm{f}} \rightarrow k_{\max} = \frac{4\pi|M^2|}{2\pi\hbar} D_{\mathrm{s}}. \tag{13}$$

Following the reasoning of Bevan *et al.*[1], we can write the expression for the current density generated by these mechanisms as:

$$J(t) = e[k_{\mathrm{f}}\bar{c}(t) - k_{\mathrm{b}}\bar{c}(t)], \tag{14}$$

where $e$ is the charge of a single electron (in first approximation, only single-electron exchanges can occur between the electrode and the reactant) and $\bar{c}(t)$ is the surface concentration of the reactant. The surface concentration of the reactant changes over time as electrochemical reactions gradually deplete the reactant molecules close to the electrodes.

*Electrophoretic transport*

The Hele–Shaw geometry makes the problem essentially one-dimensional with the electric field only dependent from the *z*-coordinate $\boldsymbol{E} = E(z)\hat{z}$. Furthermore, in our experimental conditions, the amount of charged impurities in solution is insufficient to form a double layer that affects the electric field in the bulk. As a consequence, the electric field in the cell remains uniform $E = U/h$; once a charged molecule is generated at the electrode, it is immediately subjected to a force $F_{\mathrm{E}} = eE$. The corresponding current flowing across the cell is described by the Nernst–Planck equation:

$$J_{\mathrm{N-P}} = -D\nabla c + cv + \frac{Dze}{k_{\mathrm{B}}T} cE, \tag{15}$$

where $D$ is the diffusion coefficient, $v$ is the velocity of the advection flows, and $c$ is taken as the number concentration (corresponding to the concentration of charged L-C8Azo species) so that $J_{\mathrm{N-P}}$ multiplied by $e$ has the units of current density. In our study, where the applied voltage is relatively low, we can assume that there is no electroconvection in the fluid so $v$ = 0. Furthermore, we can assume that all charged molecules generated at the electrode will relinquish their charge as soon as they come into contact with the electrode of the opposite sign. These two assumptions have two consequences: the concentration of the species remains constant in the bulk ($\nabla c = 0$), and there are always new molecules close to the electrode that can acquire charge (no depletion phenomena as in most cases in electrochemical reactions, so that $\bar{c}(t) = \bar{c}$ is constant in time). The fact that we can assume no depletion of reactants also explains why the steady-state current in our experiments can be sustained almost indefinitely. Therefore, using $E = U/h, v$ = 0 and $\nabla c = 0$, we can rewrite Equation 15 for the current density as:

$$J(U) = \frac{Dze^2}{k_B T h} c\, U. \tag{16}$$

Equation 16 makes evident that the current should be linear with the voltage following the assumptions which is in line with the observation of the trend of the current at high voltage.

*Simple current–voltage model*

Combining the two mechanisms quantified by Equations 14 and 16, considering that the generation of charges occurs at both electrodes, leads to the following expression for the current:

$$J(U) = 2e\bar{c}[k_{\mathrm{f}}(U) - k_{\mathrm{b}}(U)] + \frac{De^2}{k_{\mathrm{B}}Th}cU. \tag{16}$$

Furthermore, in first approximation, we can consider that after the application of a voltage bias the backward rate $k_b$ should rapidly go to zero. After term rearrangement and variables substitution, we arrive at[1]:

$$J(U) = 2e\bar{c}k_{\mathrm{max}}\,\mathrm{erfc}\left[(\mu - \varepsilon_0)/\sqrt{4\lambda k_{\mathrm{B}}T}\right] + \frac{De^2}{k_{\mathrm{B}}Th}cU, \tag{17}$$

Now, making the dependencies from $U$ explicit, we arrive at the following fitting function:

$$J(U) = K\,\mathrm{erfc}[A(U - U_0)] + B\,U, \tag{18}$$

Fitting of the model to the experimental data is attempted in Supplementary Fig. 15. The model clearly captures the main features of the current data, but unfortunately, it does not perform perfectly across the entire voltage range. In particular, the model partially fails to capture the trend of the data at low voltages where the tunneling contribution dominates. On the other hand, the model satisfactorily describes the behavior of toluene solutions L-C8Azo at high voltages, where the electrochemical reaction rate is saturated and the current increases mostly due to electrophoresis. In conclusion, the results indicate that the model based on the combination of charge exchange at the electrodes via tunneling and charge transport across the cell via electrophoresis provides a good first approximation of the behavior for diluted solutions of L-C8Azo in toluene.

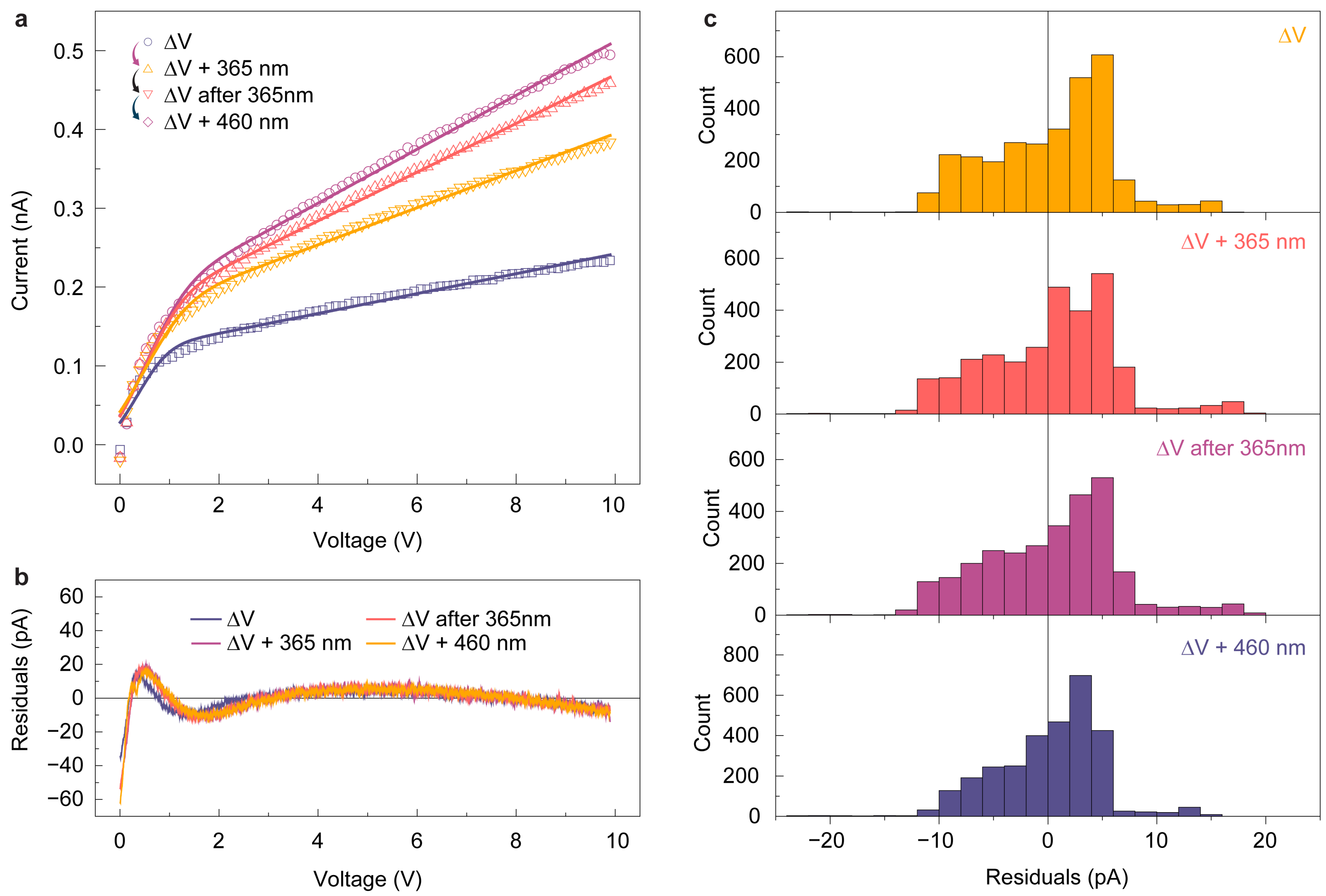

**Supplementary Fig. 15 | Testing the model describing the current–voltage data. a.** Current data for different *trans*/*cis* mixtures of L-C8Azo interpolated with Equation 18. **b.** Plot of the residuals from the interpolation of the model for the different curves. **c.** Distribution of the residuals from the interpolation of the model for the different curves.

# Supplementary Note 2: Synthesis and characterization of *para*-substituted azobenzenes

## 2.1 Materials

4,4′-Dihydroxyazobenzene (98%, ABCR), 1-bromododecane (97%, Sigma-Aldrich), 1-bromooctane (99%, Sigma-Aldrich), 1-bromo-2-ethylhexane (95%, with 1% potassium carbonate as stabilizer, Sigma-Aldrich), toluene (99%, VWR), toluene-$d_8$ (99.5% isotopic substitution, with 0.03% TMS, Thermo Scientific), *N*,*N*-dimethylformamide (extra dry, 99.8%, Thermo Scientific), iodomethane (99%, Fisher Scientific), and potassium *tert*-butoxide ( >98%, Sigma-Aldrich) were used as received.

## 2.2 Instrumentation

Analytical thin-layer chromatography (TLC) was performed on neutral aluminum sheet silica gel plates and visualized under 254 nm light. NMR spectra were recorded on a Bruker Avance IV Neo spectrometer (at 400, 600, or 800 MHz for $^{1}$H spectra, and 101, 151, or 201 MHz for $^{13}$C spectra) and processed using MestReNova (v. 15.0). $^{1}$H NMR chemical shifts in $CDCl_3$ and toluene-$d_8$ are reported in parts per million (ppm) relative to the residual solvent signals (7.26 and 2.09 ppm at 298 K, respectively). Coupling constants ($J$) are reported in Hz.

## 2.3 Synthesis of C1Azo

4,4′-dihydroxyazobenzene (200 mg, 0.934 mmol) and $K_2CO_3$ (386 mg, 2.79 mmol, 3 eq) were added to an oven-dried microwave vial under an inert atmosphere. The vial was sealed and anhydrous DMF (7 mL) was added, followed by MeI (528 mg, 232 µL, 3.72 mmol, 4 eq). The mixture was brought to 100 °C and heated at this temperature for 16 h. Upon cooling to room temperature, the reaction mixture was poured onto water (100 mL) to afford a yellow precipitate, which was collected by vacuum filtration and washed with cold water (5 × 10 mL) to afford the desired compound as a yellow solid (182 mg, 0.751 mmol, 80%). $^{1}$H NMR (400 MHz, $CDCl_3$, 298 K) δ (ppm) 7.91–7.85 (m, 4H), 7.03–6.97 (m, 4H), 3.89 (s, 6H). Spectroscopic data are consistent with the literature report[4].

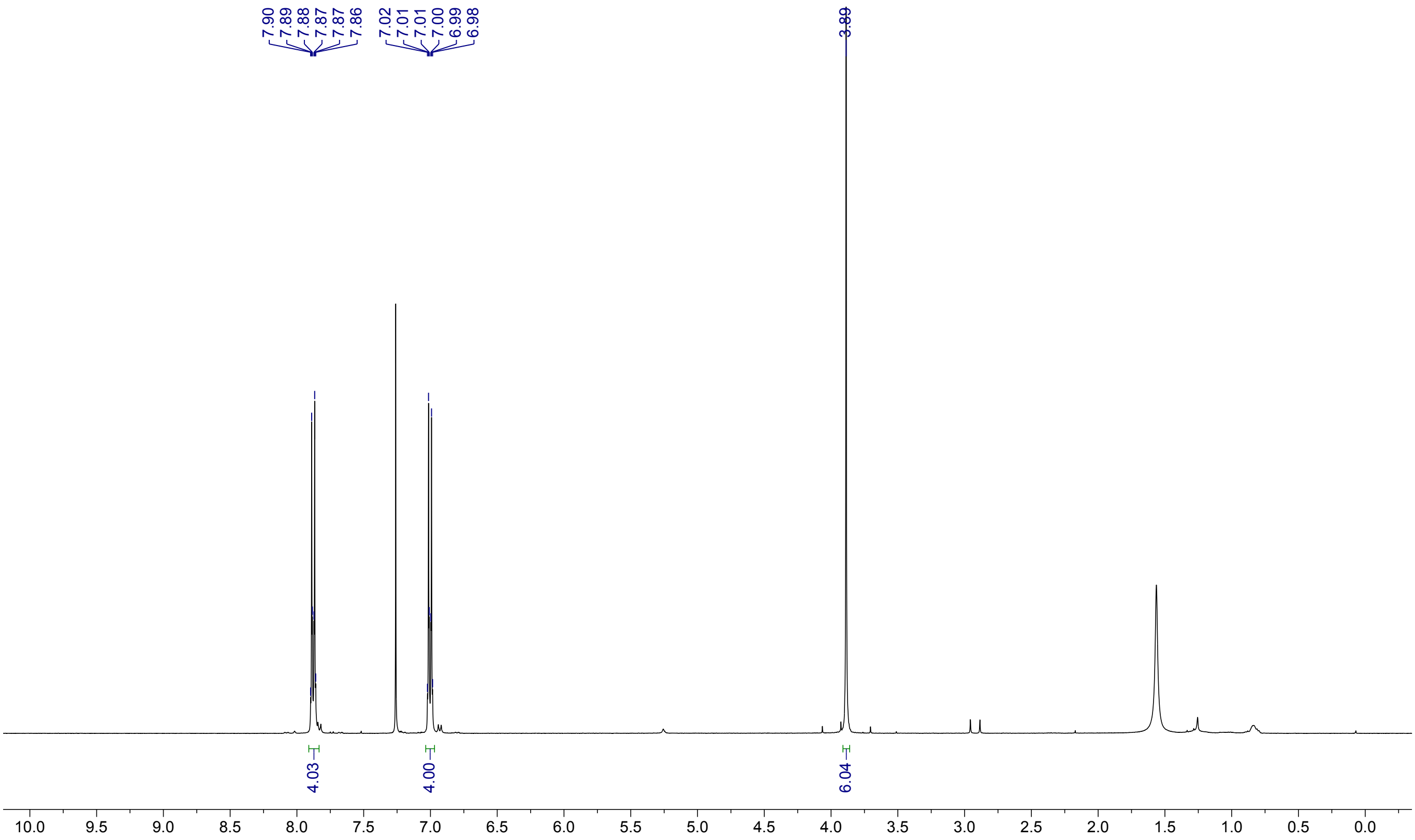

**Figure S16.** $^{1}H$ NMR spectrum of *trans*-L-C1Azo (400 MHz, $CDCl_3$, 298 K).

## 2.4 Synthesis of L-C8Azo

DMF (8 mL) was added to a mixture of 4,4′-dihydroxyazobenzene (0.40 g, 1.9 mmol), $K_2CO_3$ (0.52 g, 3.8 mmol, 2 eq), and 1-octylbromide (0.72 g, 4.2 mmol, 2.2 eq) in a round-bottom flask. The mixture was brought to 80 °C and heated at this temperature for 24 h. Upon cooling to room temperature, water and $CH_2Cl_2$ were added. The layers were separated and the crude product was extracted from the aqueous layer with $CH_2Cl_2$. The combined extracts were washed with water, then dried over anhydrous $MgSO_4$. The solvent was removed under reduced pressure, leaving a yellow residue, which was purified by column chromatography (Teledyne Isco CombiFlash Rf+ system, 40 g $SiO_2$, hexanes–EtOAc, 10→20% gradient elution) to afford the desired compound as a yellow solid (0.25 g, 0.58 mmol, 30%). $^{1}H$ NMR (400 MHz, $CDCl_3$, 298 K) δ (ppm) 7.92–7.80 (m, 4H), 7.04–6.95 (m, 4H), 4.03 (t, $J$ = 6.6 Hz, 4H), 1.87–1.76 (m, 4H), 1.52–1.43 (m, 4H), 1.42–1.25 (m, 16H), 0.95–0.84 (m, 6H).

Spectroscopic data are consistent with the literature report[5].

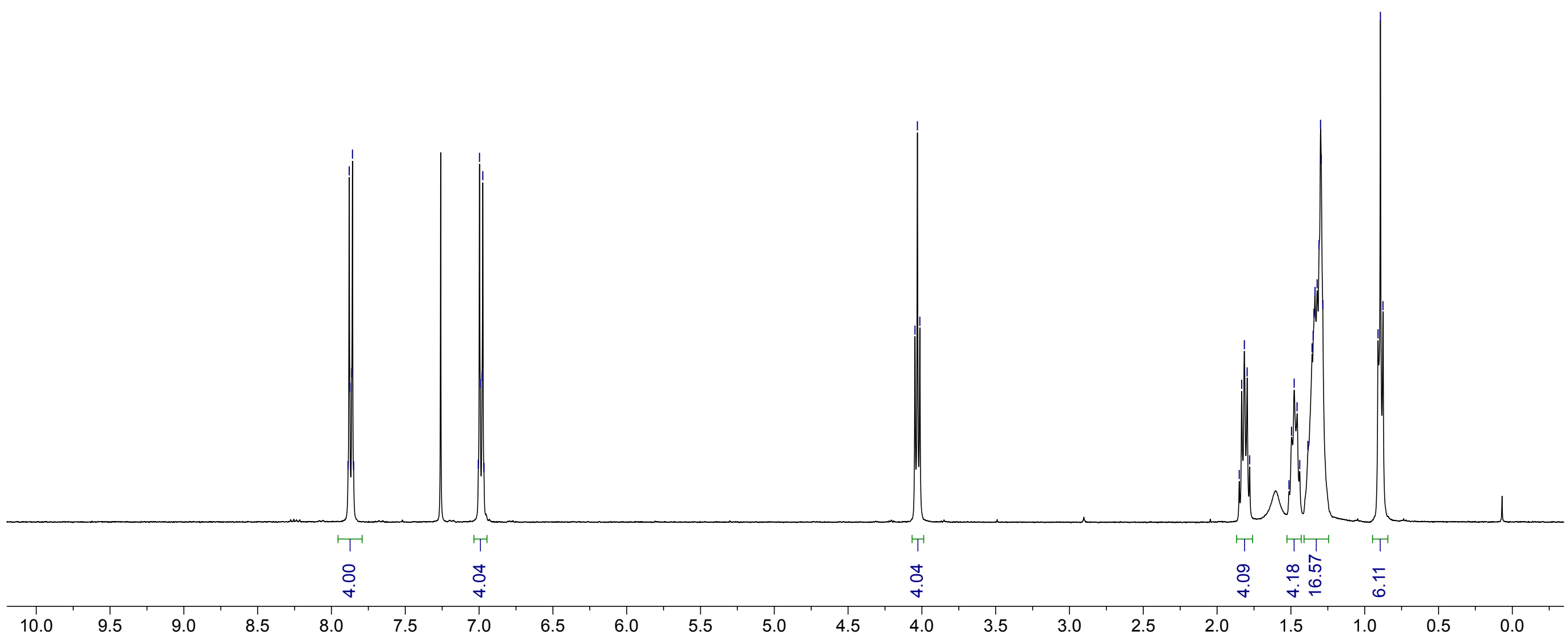

**Figure S17**. [1]H NMR spectrum of *trans*-L-C8Azo (400 MHz, $CDCl_3$, 298 K).

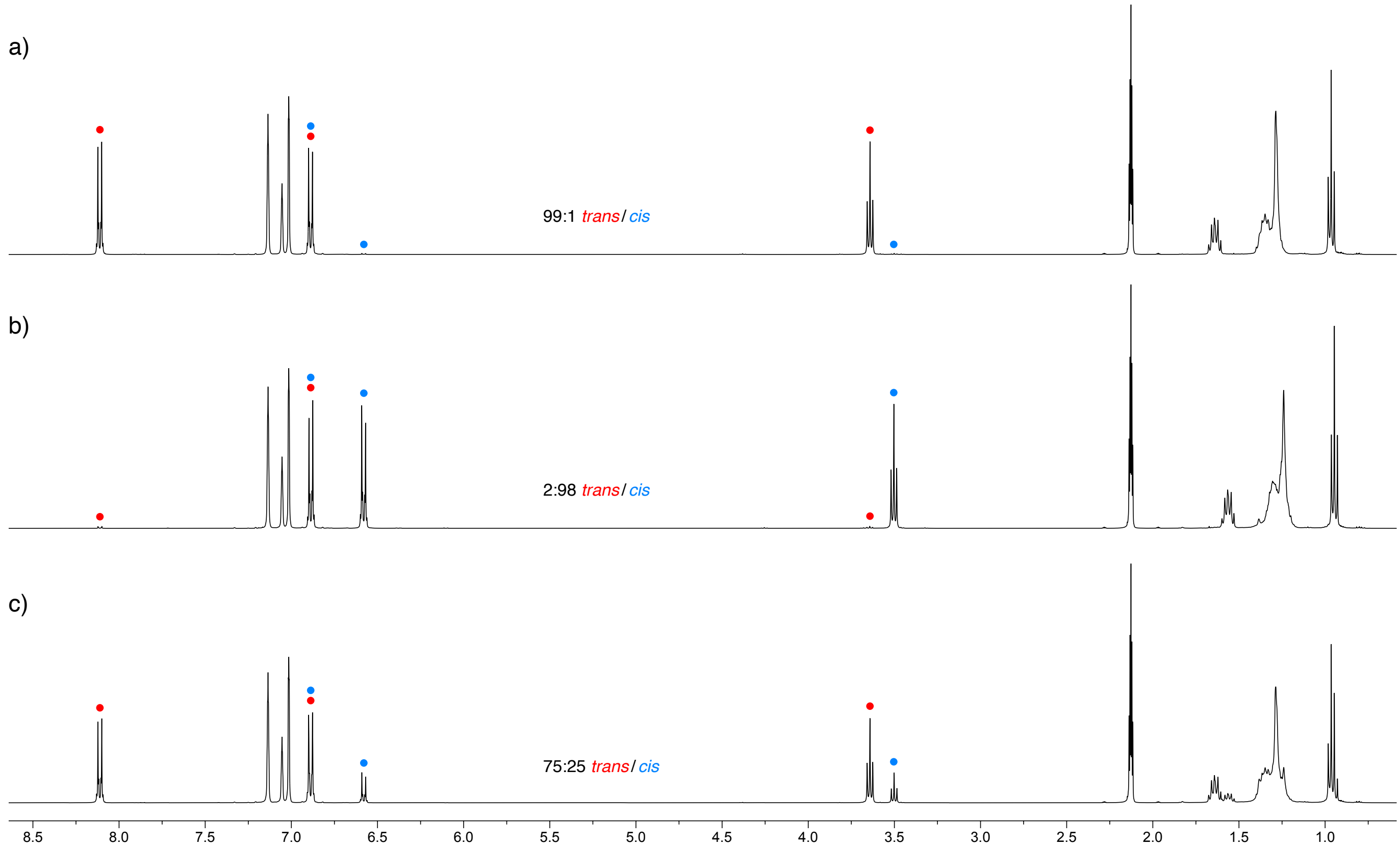

**Figure S18.** [1]H NMR spectral stack (400 MHz, toluene-$d_8$, 298 K, $c$ = 15 mM) of (a) L-C8Azo in the dark (with the initial isomeric distribution of 99:1 *trans*/*cis*); (b) after UV irradiation ($\lambda_{ex}$ = 365 nm) for 90 min (*trans*→*cis* isomerization leading to a photostationary state (PSS) that contains a 99:1 ratio of *trans*/*cis*); and (c) after the subsequent irradiation with blue light ($\lambda_{ex}$ = 460 nm) (produces a PSS that contains a 75:25 ratio of of *trans*/*cis*). Selected resonances of *trans*- and *cis*-L-C8Azo are denoted with red and blue markers, respectively.

## 2.5 Synthesis of B-C8Azo

4,4′-dihydroxyazobenzene (0.214 g, 1.00 mmol) and KO$^t$Bu (0.247 g, 2.20 mmol, 2.2 eq) were added to an oven-dried microwave vial under an inert atmosphere. The vial was sealed, and anhydrous DMF (5 mL) was added, followed by 1-bromo-2-ethylhexane (0.425 g, 2.20 mmol, 2.2 eq). The mixture was brought to 100 °C and heated at this temperature for 16 h. Upon cooling to room temperature, water and $CH_2Cl_2$ were added. The layers were separated and the crude product was extracted from the aqueous layer with $CH_2Cl_2$. The combined extracts were washed with water, then dried over anhydrous $MgSO_4$. The solvent was removed under reduced pressure, leaving a yellow residue, which was purified by column chromatography (Teledyne Isco CombiFlash Rf+ system, 40 g $SiO_2$, hexanes–EtOAc, 10→20% gradient elution) to afford the desired compound as a yellow solid (80.3 mg, 0.183 mmol, 18%). $^1$H NMR (600 MHz, $CDCl_3$, 298 K) δ (ppm) 7.89–7.83 (m, 4H), 7.02–6.97 (m, 4H), 3.95–3.89 (m, 4H), 1.76 (hept, $J$ = 6.2 Hz, 2H), 1.68–1.16 (m, 16H), 0.97–0.90 (m, 12H). $^{13}$C NMR (101 MHz, $CDCl_3$, 298 K) δ (ppm) 161.6, 147.1, 124.4, 114.8, 71.0, 39.5, 30.7, 29.2, 24.0, 23.2, 14.2, 11.3.

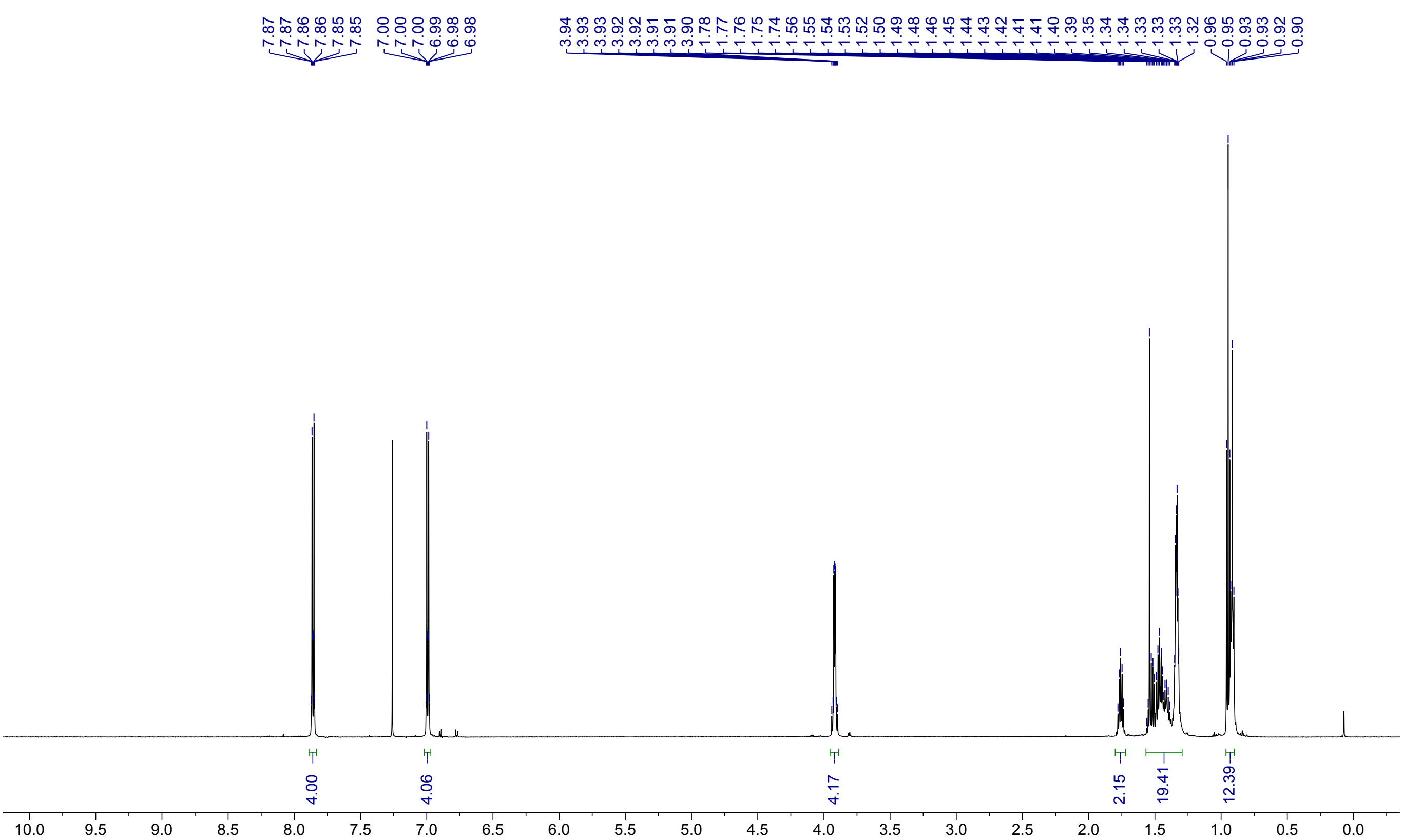

**Figure S19.** $^1$H NMR spectrum of *trans*-B-C8Azo (600 MHz, $CDCl_3$, 298 K).

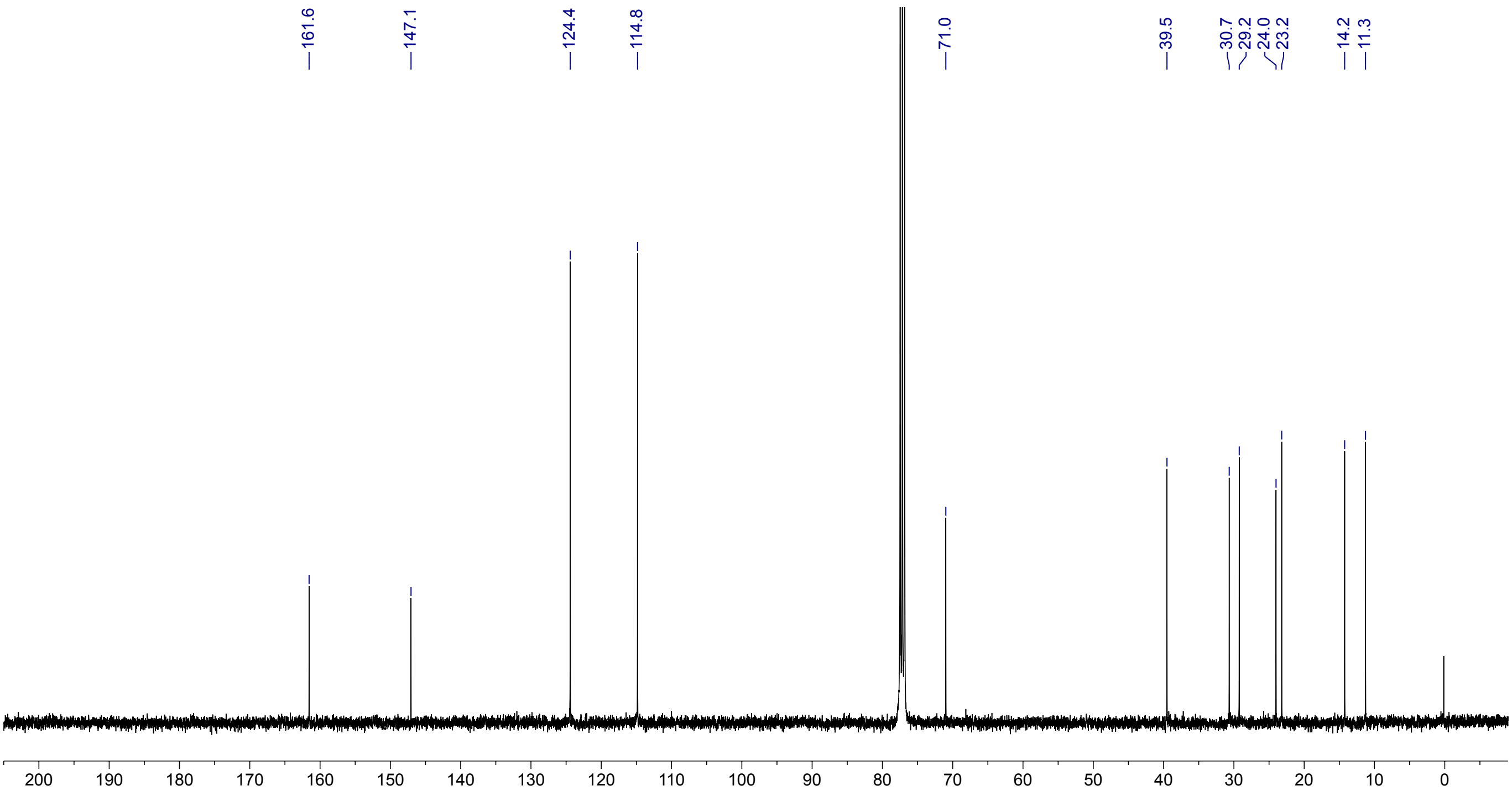

**Figure S20.** $^{13}C$ NMR spectrum of *trans*-B-C8Azo (201 MHz, $CDCl_3$, 298 K).

## 2.6 Synthesis of L-C12Azo

4,4′-dihydroxyazobenzene (0.214 g, 1.00 mmol, 1 eq) and KO$^t$Bu (0.247 g, 2.20 mmol, 2.2 eq) were added to an oven-dried microwave vial under an inert atmosphere. The vial was sealed and anhydrous DMF (5 mL) was added, followed by 1-bromododecane (0.548 g, 2.20 mmol, 2.2 eq). The mixture was brought to 100 °C and heated at this temperature for 16 h. Upon cooling to room temperature, water and $CH_2Cl_2$ were added. The layers were separated and the crude product was extracted from the aqueous layer with $CH_2Cl_2$. The combined extracts were washed with water, then dried over anhydrous $MgSO_4$. The solvent was removed under reduced pressure, leaving a yellow residue, which was purified by column chromatography (Teledyne Isco CombiFlash Rf+ system, 40 g $SiO_2$, hexanes–EtOAc, 10→20% gradient elution) to afford the desired compound as a yellow solid (47 mg, 0.086 mmol, 9%). $^1H$ NMR (600 MHz, $CDCl_3$, 298 K) δ (ppm) 7.88–7.83 (m, 4H), 7.00–6.96 (m, 4H), 4.03 (t, $J$ = 6.6 Hz, 4H), 1.84–1.79 (m, 4H), 1.49–1.45 (m, 4H), 1.40–1.22 (m, 32H), 0.88 (t, $J$ = 7.1 Hz, 6H).

Spectroscopic data are consistent with the literature report[5].

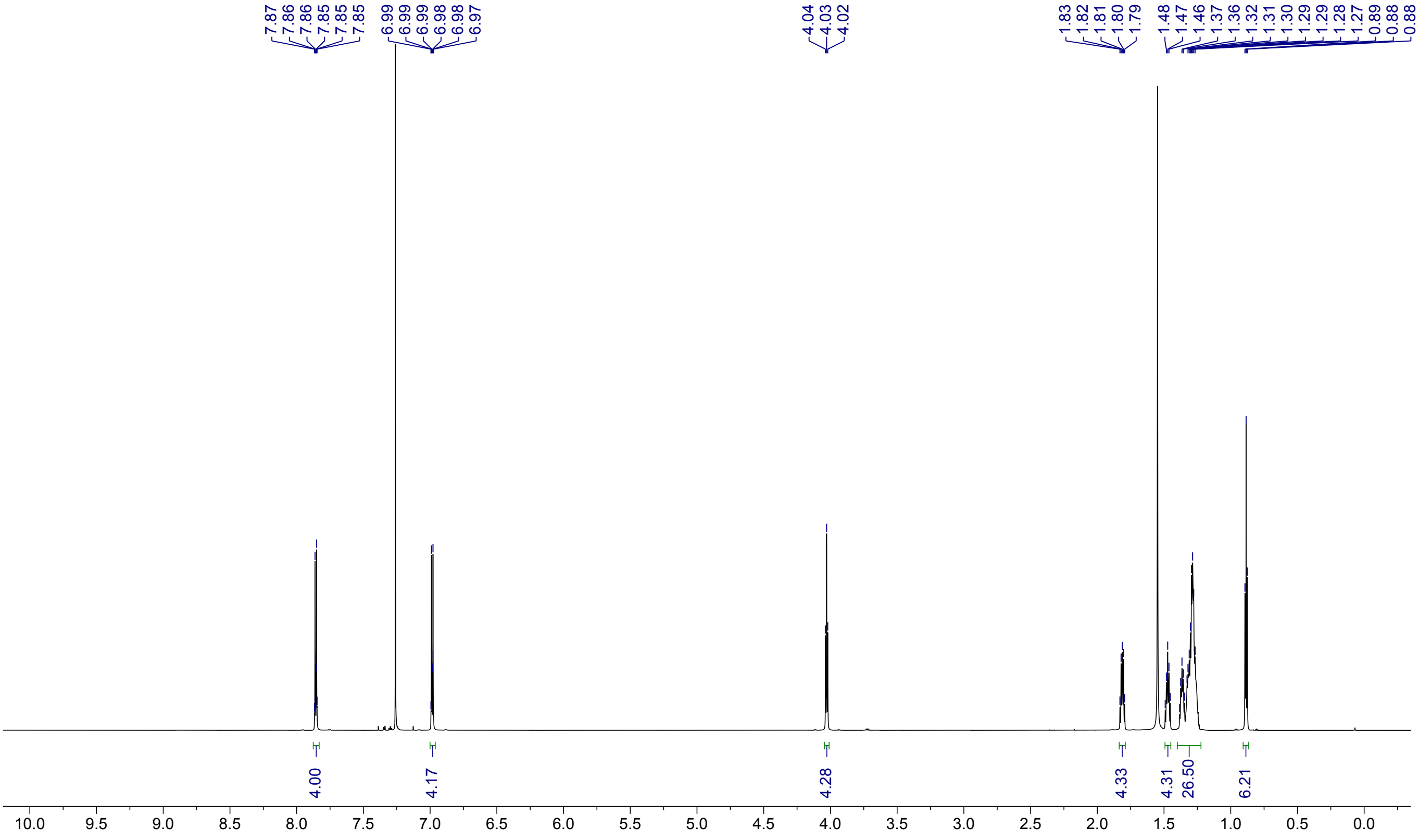

**Figure S21.** $^{1}H$ NMR spectrum of *trans*-L-C12Azo (800 MHz, $CDCl_3$, 298 K).

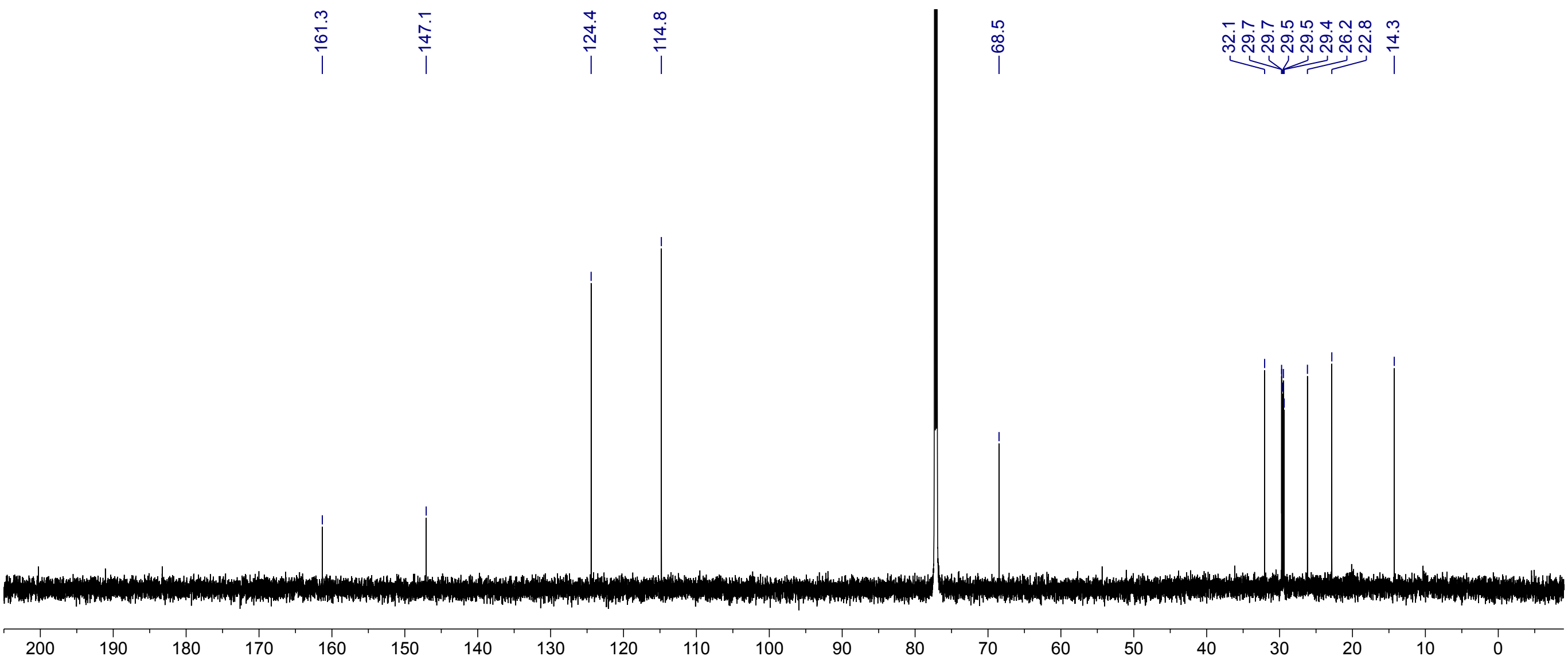

**Figure S22.** $^{13}C$ NMR spectrum of *trans*-L-C12Azo (201 MHz, $CDCl_3$, 298 K).

## References

1 Bevan, K. H., Foong, Y. W., Shirani, J., Yuan, S. & Abi Farraj, S. Physics applied to electrochemistry: Tunneling reactions. *J. Appl. Phys.* **129**, 090901 (2021).

2 Tong, X., Wang, G., Soldera, A. & Zhao, Y. How can azobenzene block copolymer vesicles be dissociated and reformed by light? *J. Phys. Chem. B.* **109**, 20281–20287 (2005).

3 Zimmerman, G., Chow, L.-Y. & Paik, U.-J. The photochemical isomerization of azobenzene. *J. Am. Chem. Soc.* **80**, 3528–3531 (1958).

4 Srivastava, A. *et al.* Light-controlled shape-changing azomacrocycles exhibiting reversible modulation of pyrene fluorescence emission. *Org. Biomol. Chem.* **20**, 5284–5292 (2022).

5 Kidowaki, M., Moriyama, M., Wada, M. & Tamaoki, N. Molecular mechanism of anomalous increase in the helical pitch of cholesteric liquid crystals induced by achiral dopants. *J. Phys. Chem. B* **107**, 12054–12061 (2003).